\documentclass{aa} %[referee]   
\usepackage{graphicx}
\usepackage{txfonts}
\usepackage{hyperref}

\usepackage{amsmath}    % Advanced maths commands
\usepackage{newtxtext,newtxmath}

\begin{document}

\title{Multiple Stellar Populations outside the tidal radius of NGC\,1851 through Gaia DR3 XP Spectra}

\author{
    Giacomo Cordoni \inst{1} \and Anna F. Marino \inst{1}$^{\and}$ \inst{2} \and Antonino P. Milone \inst{1}$^{\and}$ \inst{3} \and Emanuele Dondoglio \inst{3} \and Edoardo P. Lagioia \inst{3} \and Maria Vittoria Legnardi \inst{3} \and Anjana Mohandasan \inst{3} \and Sohee Jang \inst{4} \and Tuila Ziliotto \inst{3}
}
\institute{
    Istituto Nazionale di Astrofisica - Osservatorio Astronomico di Padova, Vicolo dell’Osservatorio 5, IT-35122, Padova, Italy
    \and 
    Istituto Nazionale di Astrofisica - Osservatorio Astrofisico di Arcetri, Largo Enrico Fermi, 5, Firenze, IT-50125, Firenze, Italy
    \and 
    Dipartimento di Fisica e Astronomia ``Galileo Galilei'', Univ. di Padova, Vicolo dell'Osservatorio 3, I-35122 Padova, Italy
    \and
    Center for Galaxy Evolution Research and Department of Astronomy,Yonsei University, Seoul 03722, Korea
}

\abstract
{}
{Ancient Galactic Globular Clusters (GCs) have long fascinated astronomers due to their intriguing multiple stellar populations (MPs) characterized by variations in light-element abundances. Among these clusters, Type-II GCs stand out as they exhibit stars with 
large differences in
%distinct 
heavy-element chemical abundances. %patterns. 
These enigmatic clusters, comprising approximately 17\% of analyzed GCs with MPs, have been hypothesized to be %associated with 
the remnants of accreted dwarf galaxies.}
{%In this study, 
We focus on 
%NGC\,1851, 
one of the most debated Type~II GCs, NGC\,1851, to investigate its
%the presence and characteristics of 
MPs across a wide spatial range of up to 50 arcmin from the cluster center. By using Gaia Data Release 3 (DR3) low-resolution XP spectra, we generate  synthetic photometry to perform a comprehensive analysis of the spatial distribution and kinematics of the canonical and anomalous populations within this GC.
By using appropriate color-magnitude diagrams (CMDs) from the synthetic photometry in the BVI bands and in the $\rm f415^{25}$ band introduced in this work, we identify distinct stellar sequences associated with different heavy-element chemical composition.
%content populations.
}
{Our results suggest that 
%We detected 
the canonical and the anomalous populations 
reside
both inside and outside the tidal radius of NGC\,1851, up to a distance that exceeds by 3.5 times its tidal radius.
%of NGC\,1851. 
%Moreover, our analysis reveals that 
However, $\sim$80\% of stars outside the tidal radius are consistent with belonging to the canonical population, emphasizing its dominance in the cluster's outer regions.
Remarkably, canonical stars exhibit a more circular on-sky morphology, while the anomalous population displays an elliptical shape. 
%Moreover, our analysis reveals that $\sim$80\% of stars outside the tidal radius are consistent with belonging to the canonical population, emphasizing its dominance in the cluster's outer regions.
Furthermore, we delve into the kinematics of the multiple populations, examining velocity dispersions, rotation patterns, and potential substructures. Our results reveal a flat/increasing velocity dispersion profile in the outer regions. Additionally, we observe hints of a tangentially anisotropic motion in the outer regions, indicating a preference for stars to escape on radial orbits.
{Our work demonstrates the capability of synthetic photometry, based on Gaia spectra, to explore multiple populations across the entire cluster field.}}
{}

\titlerunning{The Halo of NGC\,1851}
\authorrunning{Cordoni et al.}

\maketitle

\section{Introduction}
\label{sec:intro}
Numerous photometric and spectroscopic studies have established the presence of multiple stellar populations (MPs) in %ancient 
Galactic globular clusters (GCs). Type-I GCs\footnote{In this study, we will adopt the terminology introduced by \citet{milone2017}, where type-I GCs are characterized by the presence of typical multiple populations with variations in light-element abundances. On the other hand, clusters that exhibit variations in both light- and heavy-elements are classified as type-II GCs.}, which account for roughly 83\% of the 58 GCs investigated by \citet{milone2017} using multiband UV-optical Hubble Space Telescope (HST) photometry, exhibit variations in light element abundances, such as carbon, oxygen, nitrogen, and sodium. Stars with different light-elements abundances are commonly referred to as 1P/2P or 1G/2G. Conversely, type-II or anomalous GCs, which constitute the remaining 17\% of the clusters, exhibit internal variations in both light and heavy elements, such as elements associated with $s$-processes and metallicity \citep{marino2019}. Specifically, stars can be grouped into canonical and anomalous stars based on their heavy-element content, with the latter being metal-richer than the former. Despite numerous studies, the detailed formation process of MPs in GCs remains unclear, and none of the proposed scenarios can fully account for the observational constraints \citep[see][for recent reviews]{bastian2018b, milone2022a}.

One hypothesis for the origin of light-elements variations suggests that 2P stars were formed from the ejecta of more massive 1P stars after the gas collapsed in the cluster center, creating a more centrally concentrated environment. This scenario is supported by studies such as \citet{decressin2007}, \citet{dercole2010}, \citet{denissenkov2014}, and \citet{renzini2022}. Alternatively, another hypothesis suggests that GCs host only a single generation of stars, and stars with different chemical compositions result from specific exotic physical phenomena of proto-GCs, as proposed by studies such as %\citet{demink2009}, 
\citet{bastian2013}, and \citet{gieles2018}. 
Noticeably, the formation of the more complexes type-II clusters has also been associated with the nuclear remnants of accreted dwarf galaxies. The broader implication is that these stellar systems would be somehow related to the satellites accreted by the Milky Way during its formation. Indeed, as shown by \citet{milone2020}, the location of these peculiar clusters in the integral space of motions is consistent with most of them being associated with the Gaia Enceladus structure \citep{helmi2018}.
Hence, the characterization and understanding of type-II GCs would play a crucial role in the context of Milky Way formation and evolution.

Among Type-II GCs, NGC\,1851 represents one of the most investigated and debated clusters. Indeed, over the last two decades numerous literature photometric and spectroscopic works attempted at a detailed characterization of the properties of the canonical and anomalous stellar populations. However, there is still no agreement on the origin of the puzzling anomalous population. Specifically, it is not clear whether canonical and anomalous stars differ in total C+N+O abundances, $s$-elements content, metallicity and/or age \citep[see e.g.][]{milone2008, cassisi2008, ventura2009, yong2009, yong2015, carretta2010, marino2014, simpson2017, tautvaisiene2022}. 
%APM. Aggiungiamo poi questa referenza quando sara' accettato
%We refer to Dondoglio et al., (in preparation) for a detailed description of the controversial literature on NGC\,1851. 

%APM. In questo punto parlerei dell'alone. 
%  La scoperta (Olszewski et al. 2009, cita pure il paper con Da Costa, Kuzma), poi la conferma con velocita' radiali (Sollima). Ed infin il lavoro di cui parli sotto sulla caratterizzazione chimica adi Fabiola. Specifica che pero' il numero di stelle di alone analizzato e' relativamente piccolo 
%

An intriguing feature of NGC\,1851 is provided by the presence of a stellar halo encircling the cluster and extending up to more than 500 pc from its center, which was first reported by \citet{olszewski2009}, and then confirmed by several other analysis \citep{sollima2012, kuzma2018, carballo-bello, shipp18}. Additionally, some studies \citep[see e.g.][]{shipp18} have detected possible evidence of tidal tails aligned with NGC\,1851's orbit, implying potential tidal stripping due to interactions with the Milky Way.
%Furthermore, 
\citet{marino2014} utilized high-resolution spectroscopic observations and photometric data to delve into the intricate nature of NGC\,1851 and its halo. %Their comprehensive investigation yielded significant insights into the formation and origin of this unique cluster system. %Notably, the authors confirmed the presence of a stellar halo encircling NGC\,1851, extending over a span of up to 2.5 times the tidal radius. %ADD    REF FOR 
Intriguingly, through an analysis of the chemical abundances of the $s$-process elements Sr and Ba, \citet{marino2014} determined that all the seven studied  extra-tidal stars exhibited consistency with $s$-normal stars, i.e. the canonical stellar population observed within NGC\,1851. This discovery posed fascinating questions regarding the genesis of these halo stars and their connection to the primary cluster population. 

The intricate nature of the combined NGC\,1851 cluster and its surrounding halo suggests the possibility of a scenario involving a tidally-disrupted dwarf galaxy, within which NGC\,1851 was originally embedded.
A similar idea was suggested by \citet{Bekki&Yong12}, who outlined a dynamically plausible scenario with two GCs in a dwarf galaxy merging and forming a new nuclear star cluster surrounded by field stars of the host dwarf. The host dwarf galaxy is stripped through tidal interaction with the Milky Way leaving the stellar nucleus which is observed as NGC\,1851.

Regardless of the efforts in the field, the origin of the anomalous stellar population of NGC\,1851 remains concealed. Is it the result of a merging, or is it the result of peculiar star formation processes?
The answer to these questions may reside among stars beyond the tidal tail of the cluster. Indeed, in the context of the formation of MPs, the properties of the stars in the cluster outskirts may provide the smoking gun to disentangle between different formation scenarios, as they could still retain the imprint of the physical processes that shape GCs. 
%AFM 
However, despite the crucial constraints that the analysis of chemical abundances among extra-tidal stars of GCs can provide, these studies have been difficult to carry out. The main challenge is to observe a sizable number of stars in the outskirt regions that are associated with the cluster, and that are bright enough to make spectroscopic analysis of chemical abundances feasible.

In this work, we introduce a new approach to extend the study of the canonical and anomalous stellar populations in NGC\,1851, and for the first time, investigate the chemical and dynamical properties of extra-tidal stars in NGC\,1851.
Remarkably, the recent publication of Gaia Data Release 3 \citep[DR3,][]{gaiadr3} and Gaia BP/RP spectra \citep[hereafter XP spectra, ][]{montegriffo2022a,  deangeli2022}  allows us to take the study of MPs in GCs to a next level. Indeed, thanks to Gaia DR3 XP spectra, currently available for $\sim220$ million sources brighter than $G=17.5$, we have the opportunity to generate synthethic photometry in virtually any photometric systems for all the stars observed by Gaia. 
In the context of Galactic GCs, this offers an unparalleled opportunity. While HST photometry is limited to the innermost cluster regions, approximately within 1.5 arcmin, and ground-based observations typically extend up to around 15-20 arcmin from the cluster center, Gaia DR3 XP spectra enable us to explore BVI synthetic photometry across the entire cluster field. This allows for the investigation of MPs spanning from the innermost regions to the relatively unexplored tidal radii and beyond. Notably, our novel approach discussed in this study allows us to study MPs up to approximately 60 arcmin from the cluster center, thereby significantly extending our understanding of the cluster's population distribution.

% In the context of Galactic GCs, this \textbf{gives us an unprecedented opportunity}. \textbf{Indeed, while HST photometry is limited to the innermost cluster regions, i.e. within $\sim$1.5 arcmin, and available ground-based observations extend up to $\sim$15-20 arcmin from the cluster center, Gaia DR3 XP spectra allow us to }exploit UBVI synthethic photometry to investigate MPs over the whole cluster field, from the innermost regions, to the quite-unexplored clusters' tidal radii and beyond. \textbf{Specifically, we will show that the new approach discussed in this work allow to study MPs up to $\sim$ 60 arcmin from the cluster center.} 
%Moreover, Gaia XP spectra are currently available for $\sim220$ million sources brighter than $G=17.5$. 

The paper is organized as follows: in Section]\ref{sec:data} we discuss the data preparation, cluster members selection, and in Section~\ref{sec:mp} we present the clusters' Color-Magnitude Diagrams (CMDs) with their multiple populations. The analysis of the morphology and  dynamics of MPs is carried out in Section~\ref{sec:properties}. Finally, Section~\ref{sec:conclusion} provides the discussion and summary of the results.

\section{Data preparation}
\label{sec:data}

In order to build CMDs suited for the investigation of different stellar populations in type\,II GCs, we need high-quality photometry in multiple filters. %, including \textit{U}, \textit{B}, \textit{V}, and \textit{I}. 
Different filter combinations can indeed reveal the presence of MPs with distinct light and heavy element abundances. %For example, the pseudo color index $C_{\rm UBI}=(U-B)-(B-I)$, first introduced in \citet{marino2015} and widely used in MP studies, is a good indicator of the light element content of red giant branch (RGB) and main sequence (MS) stars. Due to the interplay of different molecules involving C, N, O abundances, it enables a clear separation of 1P and 2P stars in GCs \citep[e.g.,][]{milone2012, piotto2015, cordoni2020a}. 
%On the other hand, the $U-I$ color and the pseudo color 
In particular, the 
index $C_{\rm BVI}=(B-V)-(V-I)$ have been shown to effectively separate stellar populations with different heavy element and overall C$+$N$+$O content among bright RGB stars \citep{marino2015}. Hence, %these indexes are 
it is suited to investigate the properties of canonical and anomalous stars in Type-II GCs \citep[e.g.,][]{marino2015} \footnote{Various photometric diagrams constructed with ultraviolet filters, like the $B$ vs.\,$U-B$, $B$ vs.\,$U-B+I$, and $B$ vs.\,$U-2B+1$, are efficient tools to identify stellar populations with different light-element abundance in GCs \citep[e.g.][]{marino2008, milone2012, jang2022}. However, we verified that the $U$-band photometry derived from Gaia DR3 XP spectra of NGC\,1851 is not accurate enough to disentangle 1P and 2P stars. }. 

Thanks to the exquisite data provided by Gaia DR3, we are able to build different synthetic pseudo-colors to investigate MPs in GCs. In the following we describe our procedure to select bona-fide cluster members and to build appropriate CMDs in the type-II GC NGC\,1851. 
%exploit the 
Specifically, we will exploit the synthetic $C_{\rm BVI}$ pseudo color to identify and investigate canonical and anomalous stars up to $\sim 45$\, arcmin from the cluster centers, i.e. far beyond the tidal radius, namely 11.7 arcmin \footnote{Even though there is little agreement on the exact value of the NGC\,1851 tidal radius, \citep[see e.g.][revision of 2010]{mclaughlin2005, harris1996}, here we opt for the largest value.}. Hence, we will study the halo of stars surrounding NGC\,1851, and compare the properties of the different stellar populations.

\subsection{Selection of cluster members}
\label{subsec:members}

\begin{figure}
    \centering
    \includegraphics[width=0.5\textwidth, trim={0cm 2cm 0cm 2cm}, clip]{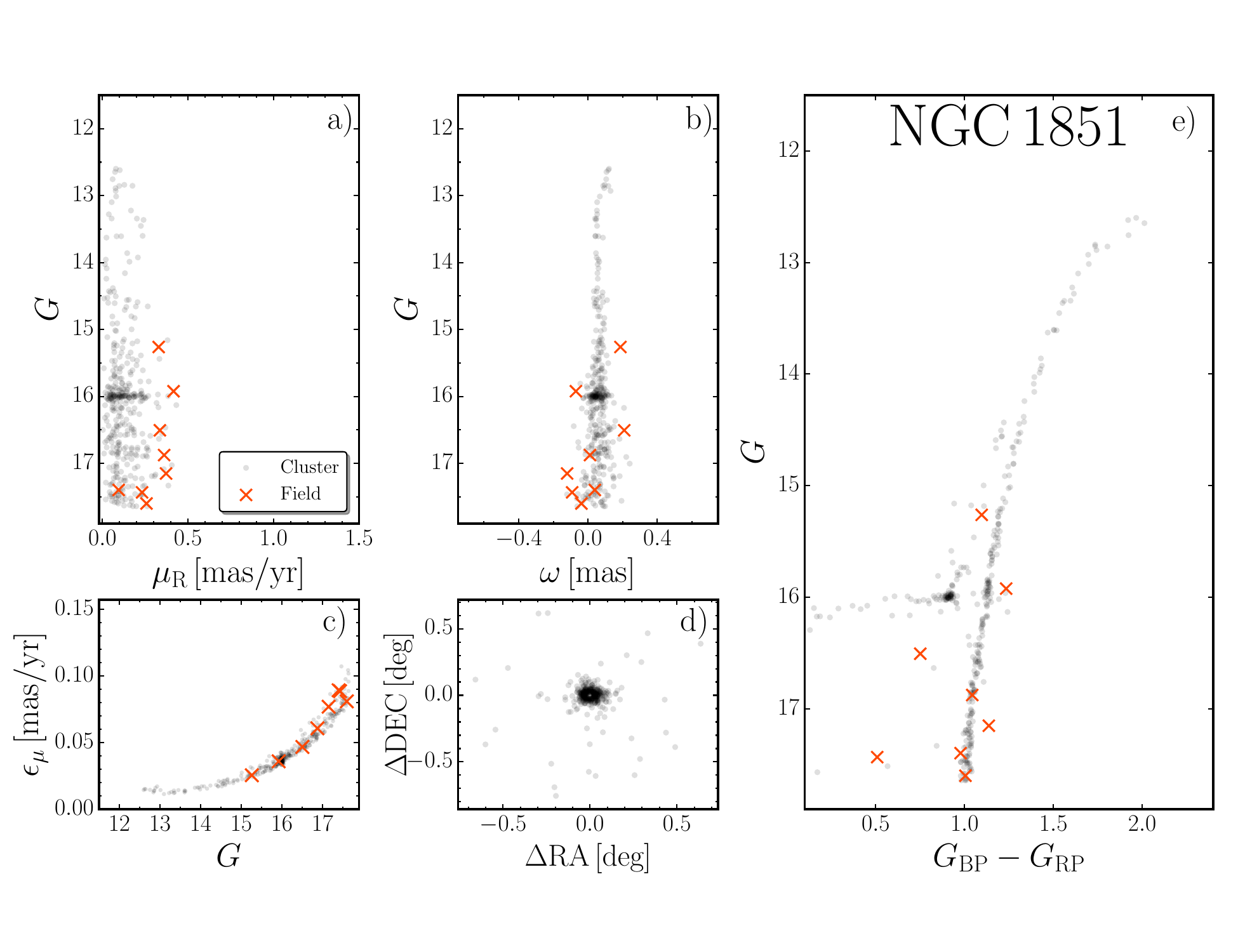}
    \caption{Cluster members selection. \textit{Panel a).} Proper motion distance $(\mu_{\rm R})$ relative to NGC\,1851 mean motion derived in \citet{vasiliev2021}, as a function of Gaia $G$ magnitude. Cluster members are displayed as black shaded points, while residual field stars contaminants are shown with orange crosses. \textit{Panel b).} Parallax $(\omega)$ as a function of $G$ magnitude. \textit{Panel c).} Uncertainty in proper motions $(\epsilon_\mu)$ vs. Gaia $G$ magnitude. \textit{Panel d).} $\Delta\,RA$ vs. $\Delta\, DEC$ of the selected cluster members. \textit{Panel e).} $G$ vs. $G_{\rm BP} - G_{\rm RP}$ CMD of the selected cluster members.} 
    \label{fig:sel}
\end{figure}

To investigate the properties and kinematics of MPs in GCs we need to \textit{i)} identify stars with accurate astrometric and photometric measurements and \textit{ii)} disentangle cluster members from field stars contaminants. To this goal, we adapted the method discussed in \citet{milone2018, cordoni2020a, cordoni2020b, cordoni2023}. The results of the selection process are shown in Fig.~\ref{fig:sel}  for the GC NGC\,1851. Moreover, we exploited different Gaia DR3 photometric and astrometric parameters, and the quality selection criteria discussed in \citet{riello2018, riello2021, montegriffo2022b}. The adopted quality criteria are the following: \textit{i)} \texttt{RUWE}<1.4, \textit{ii)} \texttt{IPD\_frac\_multi\_peak}<7, \textit{iii)} \texttt{IPD\_frac\_odd\_win}<7, \textit{iv)} $|C_{*}|<\sigma_{\rm C*}$ \citep[see e.g. Section 9.2 in][]{riello2021}, \textit{v)} $\epsilon_{\rm \mu}<0.1\rm \,mas/yr$, where $\epsilon_{\mu}$ is defined in \citet{lindegren2018}
\begin{align}
\epsilon_{\rm \mu} &= \sqrt{\frac{1}{2}\left(C_{\rm 33} +  C_{\rm 44}\right) + \frac{1}{2}\sqrt{\left(C_{\rm 44} - C_{\rm 33} \right)^2 + 4C_{\rm 34}^2}} \\
C_{\rm 33} &= \epsilon_{\mu_{RA*}}^2 \\
C_{\rm 34} &= \epsilon_{\mu_{RA*}}\epsilon_{\mu_{DEC}}\rho \\
C_{\rm 44} &= \epsilon_{\mu_{DEC}}^2
\end{align}
and is shown in Fig.~\ref{fig:sel}c. We applied constraints \textit{ii)} and \textit{iii)} to eliminate sources with multiple peaks (\texttt{IPD\_frac\_multi\_peak}) or significant contamination from nearby sources (\texttt{IPD\_frac\_odd\_win}), both detected in over 7\% of the total number of transits. For detailed information about these parameters, please refer to the Gaia DR3 documentation\footnote{\url{https://gea.esac.esa.int/archive/documentation/GDR3/Gaia_archive/chap_datamodel/sec_dm_main_source_catalogue/ssec_dm_gaia_source.html}}. Additionally, constraints \textit{iv)} and \textit{v)} were applied to exclude sources with BP/RP flux excess and with large proper motion uncertainties.
As a consequence of the heavy crowding of the innermost regions, stars within $\sim$ 1.5 arcmin are excluded from the final sample of well measured stars. Additionally, we remind the reader that Gaia DR3 XP spectra are currently available for stars brighter than $G=17.5$, thus limiting our analysis to RGB stars.
%Finally, the selected stars and the resulting CMD of selected cluster members are represented in Fig.~\ref{fig:sel}. Specifically, Fig.~\ref{fig:sel}a shows the proper motion distance, i.e. $\mu_{\rm R}=\sqrt{(\mu_{\rm RA*}-\mu_{\rm RA*, cl})^2 + (\mu_{\rm DEC}-\mu_{\rm DEC, cl})^2}$, as a function of $G$ magnitude. 

Cluster membership of each star was determined following the procedure described in \citet{cordoni2020a, cordoni2020b}, which relies on stellar positions, proper motions and parallaxes. Briefly, we analyzed the proper-motion vector-point diagram, and we calculated the proper motion distance of each star relative to the cluster mean motion, derived in \citet{vasiliev2021}. Such distance is denoted $\mu_{\rm R}$, and is shown in Fig.~\ref{fig:sel}a as a function of Gaia $G$ magnitude. Only stars with $\mu_{\rm R}$ within 4$\sigma$ from the mean trend are retained as cluster members. The same procedure has been repeated for the parallax (Fig.~\ref{fig:sel}b).
The results obtained with the described procedure have been compared with those obtained using the approach discussed in \citet{vasiliev2021}, which is based on Gaussian Mixture Models\footnote{The code to use such routine is publicly available on GitHub \url{https://github.com/GalacticDynamics-Oxford/GaiaTools}}, finding consistent results. The position and CMD of the selected cluster members are represented in Fig.~\ref{fig:sel}d and e. 

Finally, to investigate the influence of residual field stars contamination, we applied the discussed selection criteria to an outer field with the same area as the cluster field but virtually void of cluster members. Such field is located approximately at a distance of 120 arcmin from NGC\,1851 center. Stars that fulfill the selection criteria, i.e. quality cuts, proper motions and parallaxes selection, are displayed as green crosses in Fig.~\ref{fig:sel}. A visual inspection of Fig.~\ref{fig:sel} reveals that eight field stars satisfy our membership criteria, with only 4 of them exhibiting colors and magnitudes consistent with the CMD of NGC\,1851. Furthermore, as we will discuss in Sect.\ref{sec:mp}, we restrict the analysis of multiple populations to bright RGB stars, i.e. $I<16$, thus further reducing the contribution of residual field stars contamination (three out of seven field stars have magnitudes in the analyzed range). These results clearly demonstrate the negligible effect of residual field stars contamination.

\subsection{Gaia XP spectra}
\label{subsec:xp spectra}
Recently, Gaia DR3 released low-resolution BP/RP or XP spectra with wide wavelength coverage (330 to 1050\,nm) for $\sim$ 219 million stars with $G$ magnitude brighter than 17.5.
The low-resolution spectra are released through the Gaia archive in the form of two sets of 55 Hermite function coefficients, one for each channel (BP/RP), and can be converted into standard calibrated spectra, i.e. flux vs. wavelength, by means of the public Python library \texttt{GaiaXPy}\footnote{\url{https://gaia-dpci.github.io/GaiaXPy-website/}}. Furthermore, synthethic photometry in different photometric systems can be automatically generated from the input spectra coefficients. We refer to \citet{deangeli2022, montegriffo2022a, montegriffo2022b, weiler2023} for a detailed description of Gaia XP spectra.\\
Specifically, for the purpose of this work, we generate synthetic photometry in the standardized Johnson-Kron-Cousin’s photometric system (hereafter JKC), which contains the UBVI filters, fundamental for the investigation of MPs. To validate the accuracy of our conversion, we compare our results with available Stetson photometry \citep[][]{stetson2019}.

% \subsubsection{Synthethic photometry with Gaia XP spectra}
% \label{subsec:cmds}

Following the methodology described in \citet{montegriffo2022b}, we made use of the \texttt{GaiaXPy} python package to transform Gaia DR3 XP spectra into the JKC synthetic standardized photometric system, allowing us to convert the spectra into $UBVI$ magnitudes. For more information about the photometric system, we refer to \citet{montegriffo2022b}, Section 3.2. Gaia XP spectra for approximately 220 million sources with $G<17.5$ are available and can be identified in the Gaia archive by the flag \texttt{has\_xp\_continuous=True}.

\begin{figure}
    \centering
    \includegraphics[width=0.4\textwidth, trim={0cm 0cm 1cm 0cm}, clip]{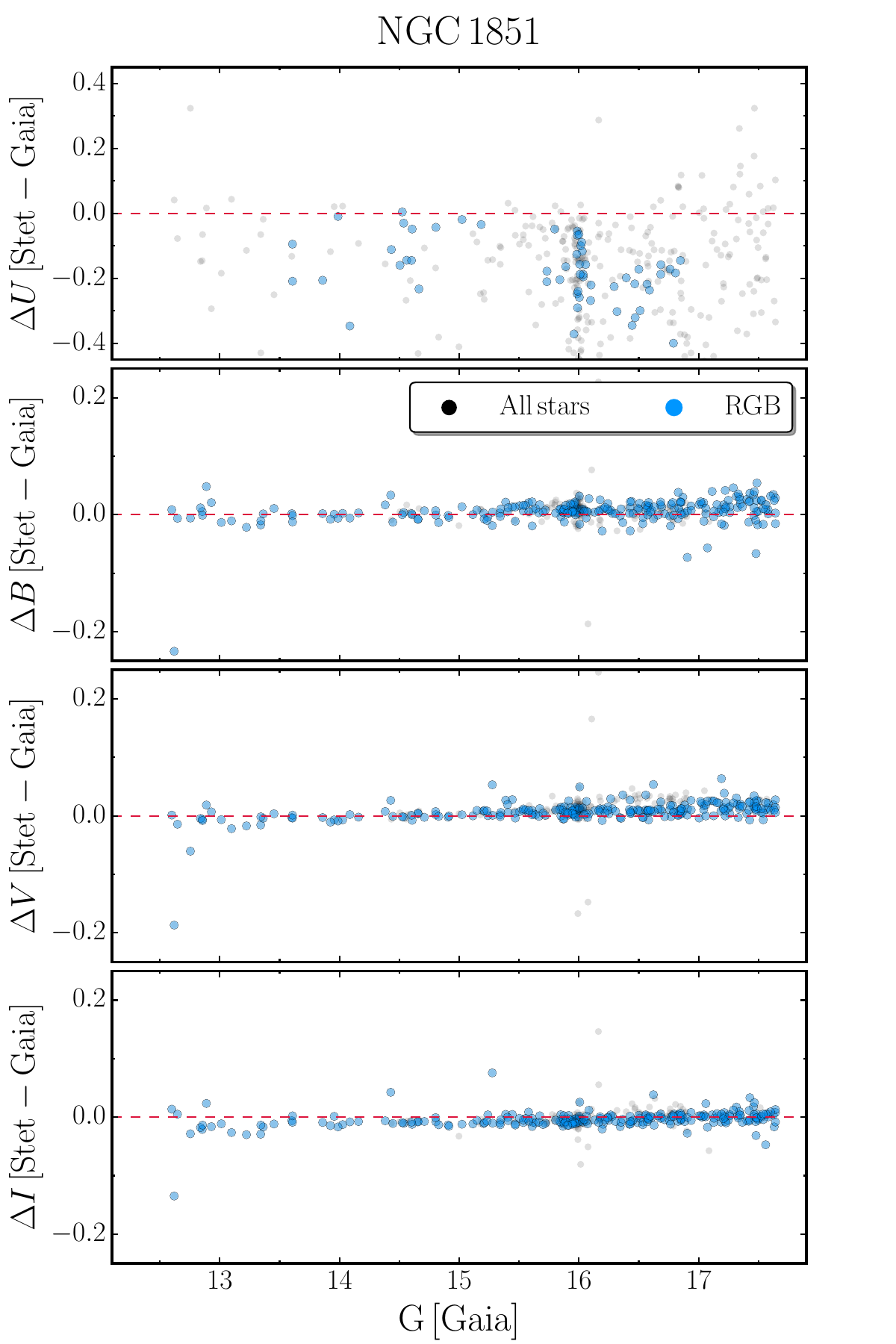}
    \caption{Gaia synthethic photometry. Comparison between Gaia synthethic photometry generated from XP spectra and the photometry from \citet{stetson2019} in the \textit{UBVI} photometric bands, from top to bottom respectively, for NGC\,1851. Gray shaded points indicate all common stars in the two catalogs, while azure dots represent RGB cluster members in NGC\,1851.}
    \label{fig:comparison}
\end{figure}

We used the \textit{UBVI} photometric bands to detect MPs among RGB stars in NGC\,1851, and verified the accuracy of generated synthetic photometry by comparing magnitudes in each band with the corresponding magnitudes provided in the Stetson's catalog \citep{stetson2019}. Fig.~\ref{fig:comparison} illustrates the comparison for stars in common between the Gaia and Stetson datasets, where $\Delta X = X_{\rm Stet}-X_{\rm Gaia}$ is plotted against $G$ for each photometric band. All common stars are represented by black points, while cluster members RGB stars are shown in azure.

The comparison reveals that synthetic \textit{BVI} magnitudes are consistent with Stetson photometry, with differences less than 0.05 magnitudes for stars brighter than $G=17$, possibly due to small variations in the photometric zero point. However, synthetic \textit{U} photometry displays larger and more variable differences with respect to Stetson \textit{U} magnitudes, likely because the \textit{U} filter is positioned at the blue edge of Gaia XP spectra, making the conversion process less reliable. Indeed, the top panel of Fig~\ref{fig:comparison} reveals a larger and random scatter with respect to that observed in the BVI filters.\footnote{We note that the \textit{y}-axis range in the top panel of Fig~\ref{fig:comparison} is almost twice as large as that displayed in all the three lower panels.} We therefore only considered stars with signal-to-noise ratio (SNR) greater than 30 when dealing with \textit{U} magnitudes, following the guidelines of \citet{montegriffo2022b}. Nevertheless, the discrepancies in \textit{U} magnitudes in NGC\,1851 are not correlated with magnitude, position or distance from the cluster center, making it impossible to use this photometric band to build the $C_{\rm UBI}$ index. However, as NGC\,1851 exhibits variations in heavy elements, we can make use of the Gaia-generated \textit{BVI} magnitudes to study the properties of canonical and anomalous stars. In fact, as shown by \citet[see e.g. their Figure 13]{marino2015}, the pseudo-color $C_{\rm BVI}=(B-V)-(V-I)$ index is an effective tool for identifying anomalous stars in Type-II GCs and investigating their properties. \\
Furthermore, we calculated the photometric uncertainties in each band using the SNR provided by the \texttt{GaiaXPy} python library. By focusing on the sample of RGB stars, we determined the average photometric uncertainties to be 0.06, 0.03, and 0.02, with dispersions of 0.02, 0.01, and 0.009, respectively, for the \textit{B, V, I} photometric filters.

\section{Multiple populations along the color-magnitude diagrams}
\label{sec:mp}
\begin{figure*}
    \centering
    \includegraphics[width=0.99\textwidth, trim={0cm 0cm 0cm 0cm}, clip]{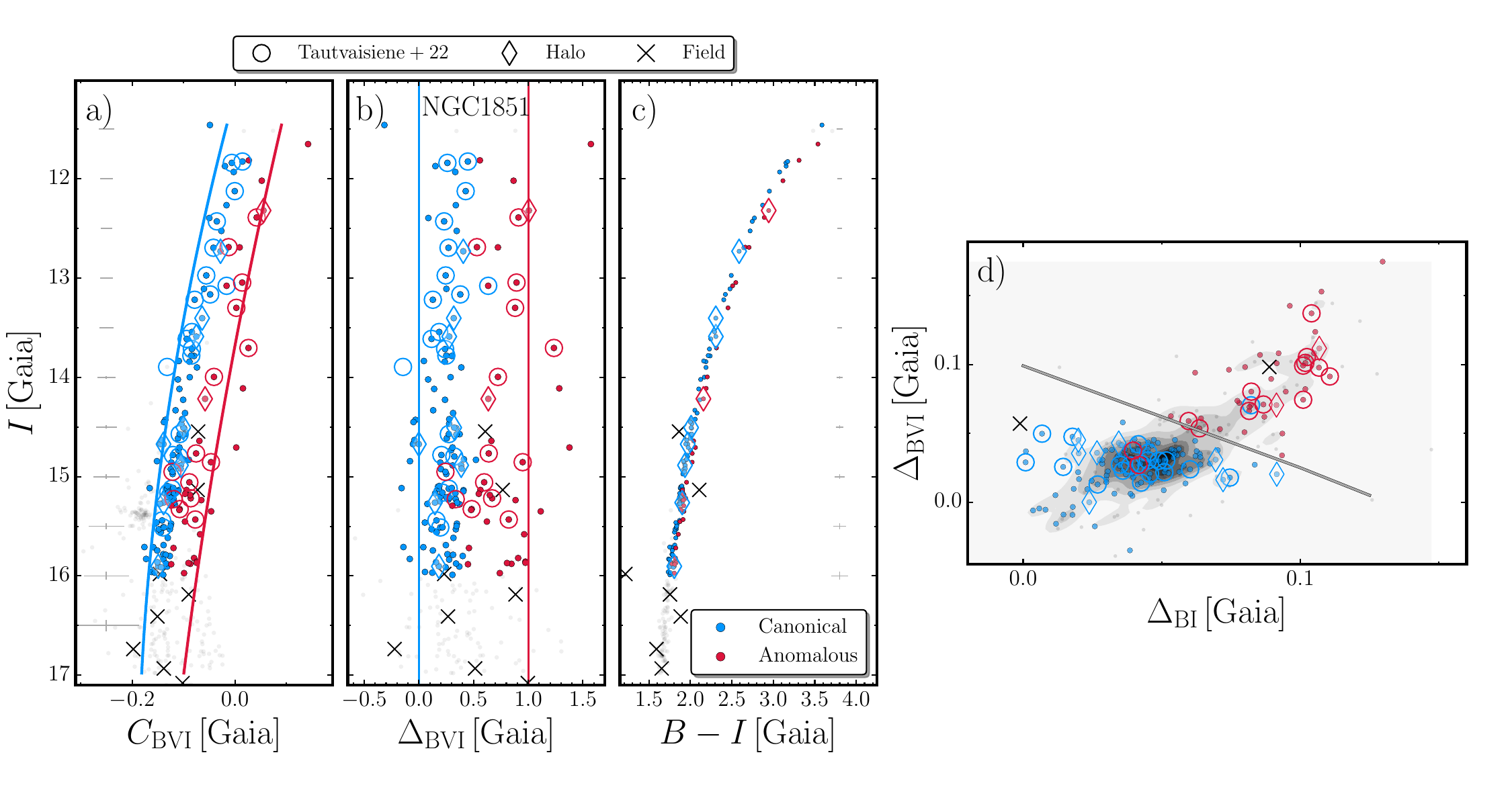}
    \caption{Multiple stellar populations along the CMD.  \textit{Panel a).} $I$ vs. $C_{\rm BVI}$ pseudo-color diagram of NGC\,1851 cluster members. Canonical and anomalous stars, selected in the Chm displayed inf panel d), are respectively marked with filled azure and red dots. Empty diamonds indicate halo stars, i.e. stars outside the 11.7 arcmin tidal radius. Azure and red empty circles indicate metal-poor and metal-rich stars identified in \citet{tautvaisiene2022}, respectively. Average photometric uncertainties are shown as gray errorbars in panel a-c. \textit{Panel b).} Verticalized $I$ vs. $\Delta_{\rm BVI}$ pseudo-color diagram color-coded as in panel a. \textit{Panel c).} $I$ vs. $B-I$ CMD. \textit{Panel d).} $\Delta_{BVI}$ vs. $\Delta_{\rm BI}$ ChM. Stars are color-coded as in panel a. The gray line represents the line used to separate canonical and anomalous stars. Gray contours in the background represent the star density distribution as determined by means of 2D Gaussian KDE with fixed bandwidth. In the whole field of view, we identified 108 and 39 candidate canonical and anomalous stars, respectively, while outside the tidal radius we found 8 and 2 canonical and anomalous stars. }
    \label{fig:popsel}
\end{figure*}

% To identify multiple stellar populations with different heavy element content in NGC\,1851, we followed the procedure described in \citet{marinoXXXX, milone2016, cordoni2020b}. Briefly, the procedure is based on the analysis of the \textit{I} vs. $C_{\rm BVI}$ and \textit{I} vs. \textit{B-I} CMD, and the resulting pseudo two-color diagram dubbed Chromosome Map (ChM). The procedure is illustrated in Fig.~\ref{fig:popsel}. The azure and red lines shown in Fig.~\ref{fig:popsel}a1-b2 represent the RGB boundaries computed as in \citet{milone2017}, i.e. as the $4^{\rm th}$ and $96^{\rm th}$ percentile of the $C_{\rm BVI}$ and $B-I$  distribution. The color of each star is then converted into the verticalized color, defined as $\Delta C=\frac{c-c_{\rm R}}{c_{\rm B}-c_{\rm R}}$ where $c, c_{\rm R}, c_{\rm B}$ are the $C_{\rm BVI}$ of the stars and of the azure and red fiducial lines, respectively. The verticalized colors are shown in Fig.~\ref{fig:popsel}a2-b2. \\
To identify multiple stellar populations with different heavy element content in NGC\,1851, we employed the methodology outlined in previous studies \citep{marino2015, milone2017, cordoni2020b}. In brief, this technique involves analyzing the \textit{I} vs. $C_{\rm BVI}$ and \textit{I} vs. \textit{B-I} CMDs, along with the resulting pseudo two-color diagram known as the Chromosome Map \citep[ChM,][]{milone2017, jang2022}. Fig.~\ref{fig:popsel} provides a visual representation of this process.

First, we determined the RGB boundaries in the as the  $4^{\rm th}$ and $96^{\rm th}$ percentiles of the $C_{\rm BVI}$ and $B-I$ distribution using the approach from \citet{milone2017}. The color of each star was then converted into the verticalized color, denoted as $\Delta C=\frac{c-c_{\rm R}}{c_{\rm B}-c_{\rm R}}$, where $c$, $c_{\rm R}$, and $c_{\rm B}$ are the $C_{\rm BVI}$ values for the stars and the blue and red fiducial lines, respectively. As a result, stars lying on the blue/red fiducial line have a $\Delta C=0$ and $1$, respectively. The resulting verticalized colors are shown in Fig.\ref{fig:popsel}b. Finally, we normalized each distribution to the color-width of the CMD at a $I$ magnitude of 15. The normalized $\Delta C$ are displayed in the ChM in Fig.~\ref{fig:popsel}d. Average photometric uncertainties are displayed in Fig.~\ref{fig:popsel}a-c) with gray errorbars.

Clearly, the $\Delta_{\rm BVI}$ vs. $\Delta_{\rm BI}$ ChM allows us to distinguish between the canonical and anomalous populations, which manifest as two distinct clumps. Specifically, considering the average photometic uncertainties and the color distribution in the $C_{\rm BVI}$ index, we restricted our analysis to stars brighter than $I=16$, for which the separation between canonical and anomalous stars is more evident. \\
To separate the two populations, we employed a combination of visual inspection and 2D Gaussian Kernel Density Estimates (KDE), shown as gray-shaded contours in the background of Fig. 3d in Fig.~\ref{fig:popsel}d. The resulting classification is displayed in Fig.~\ref{fig:popsel}d, where azure and red points correspond to canonical and anomalous stars, respectively, while the gray line indicates the adopted separation.
% Additionally, we investigated the distribution of $\delta C_{\rm BVI}$ and utilized Gaussian Mixture Models to distinguish between canonical and anomalous stars in the $I$ vs. $C_{\rm BVI}$ diagram. The best-fit Gaussians for the two populations are overlaid in Fig.~\ref{fig:popsel}c1, with azure and red colors, while the overall kernel density estimates of the distribution are represented in yellowish colors.

% To distinguish between the first and second population stars, we defined stars with $\Delta C_{\rm BVI} \leq \rm X$ as first population stars, and those with $\Delta C_{\rm BVI} > \rm X$ as second population stars. Additionally, we repeated the selection process using a smooth boundary between the first and second population stars. In the former case, we utilized the best-fit Gaussians to determine the probability of each star belonging to either the first or second population, allowing for an overlap between the two populations in $\Delta C_{\rm BVI}$. The results of both procedures were found to be consistent with each other.

% CHM --> First, we obtained the verticalized colors and utilized them to construct a pseudo two-color diagram, known as the Chromosome Map (ChM), which plots $\Delta_{\rm BVI}$ against $\Delta_{\rm BI}$.

Our analysis suggests that 108 stars are likely canonical, while the remaining 39 are more consistent with being anomalous stars. Based on this classification, the population ratio of canonical stars with respect to the total number of stars is $N_{\rm can}/N_{\rm tot}=0.735 \pm 0.036$. Specifically, 100 and 37 canonical and anomalous stars lie within the tidal radius yielding a ratio $N_{\rm Can}/N_{\rm Tot} = 0.730 \pm 0.038 $, consistent with previous HST studies \citep[see e.g.][]{milone2017}. On the other hand, 8 and 2 stars, marked with large empty azure and red diamonds in Fig.~\ref{fig:popsel}, are consistent with being extra-tidal canonical and anomalous stars, respectively ($N_{\rm Can}/N_{\rm Tot} = 0.80 \pm 0.10 $).   \\

In order to demonstrate the effectiveness of the $C_{\rm BVI}$ index in distinguishing between canonical and anomalous stars, we cross-matched our catalogs with stars with available spectroscopic information from \citet{tautvaisiene2022}. The empty large azure and red circles depicted in Fig.~\ref{fig:popsel} correspond respectively to stars classified as metal-poor and metal-rich, as outlined in their Figure 8 and Tables 1 and 2 of \citet{tautvaisiene2022}. Notably, the classification obtained using the $C_{\rm BVI}$ index is in agreement with the findings of \citet[][see Section 3 for details]{tautvaisiene2022}, with only 3 out of 35 stars exhibiting inconsistencies. \\
%As a result, we will proceed with the classification derived from the $C_{\rm BVI}$ index for the remainder of our analysis.
Additionally, we show the BVI synthethic photometry generated for the residual field stars contaminants, identified as discussed in Sec.~\ref{subsec:members}, as black crosses in Fig.~\ref{fig:popsel}. A visual inspection of Fig.~\ref{fig:popsel}a-d reveals that only two field stars exhibit pseudo-colors and magnitudes consistent with the bulk of NGC\,1851 stars. Specifically, considering the ChM in Fig.~\ref{fig:popsel}d, one star overlaps with the location of the canonical population, and one with the anomalous one. Such result further confirms the soundness of our findings concerning the canonical and anomalous populations.

\subsection{Defining an optimal filter to identify multiple populations in Type-II globular clusters}
\label{sec:custom spectra}
In the following, we will try to exploit Gaia XP spectra to build ad-hoc photometry to reflect the chemical composition of multiple stellar populations with different heavy-element content. Specifically, we will adopt the classification of Sect.~\ref{sec:mp} as a benchmark sample of canonical and anomalous stars.  Detailed analysis of synthethic spectra carried out in Dondoglio et al., in preparation, revealed that the effectiveness of the $C_{\rm BVI}$ index is mostly due to differences in the spectra of canonical and anomalous stars in the spectral range covered by the $B$ filters. Specifically, to reproduce the observations, anomalous stars must be enhanced in $C+N+O$. Specifically, as found in \citet{yong2015}, $N$ varies, while $C$ and $O$ remain constant.  On this note, we perform an exploitative analysis on canonical and anomalous XP spectra to confirm the presence of distinctive features in the $B$ spectral range. \\

\begin{figure*}
    \centering
    \includegraphics[width=0.99\textwidth, trim={0cm 0cm 0cm 0cm}, clip]{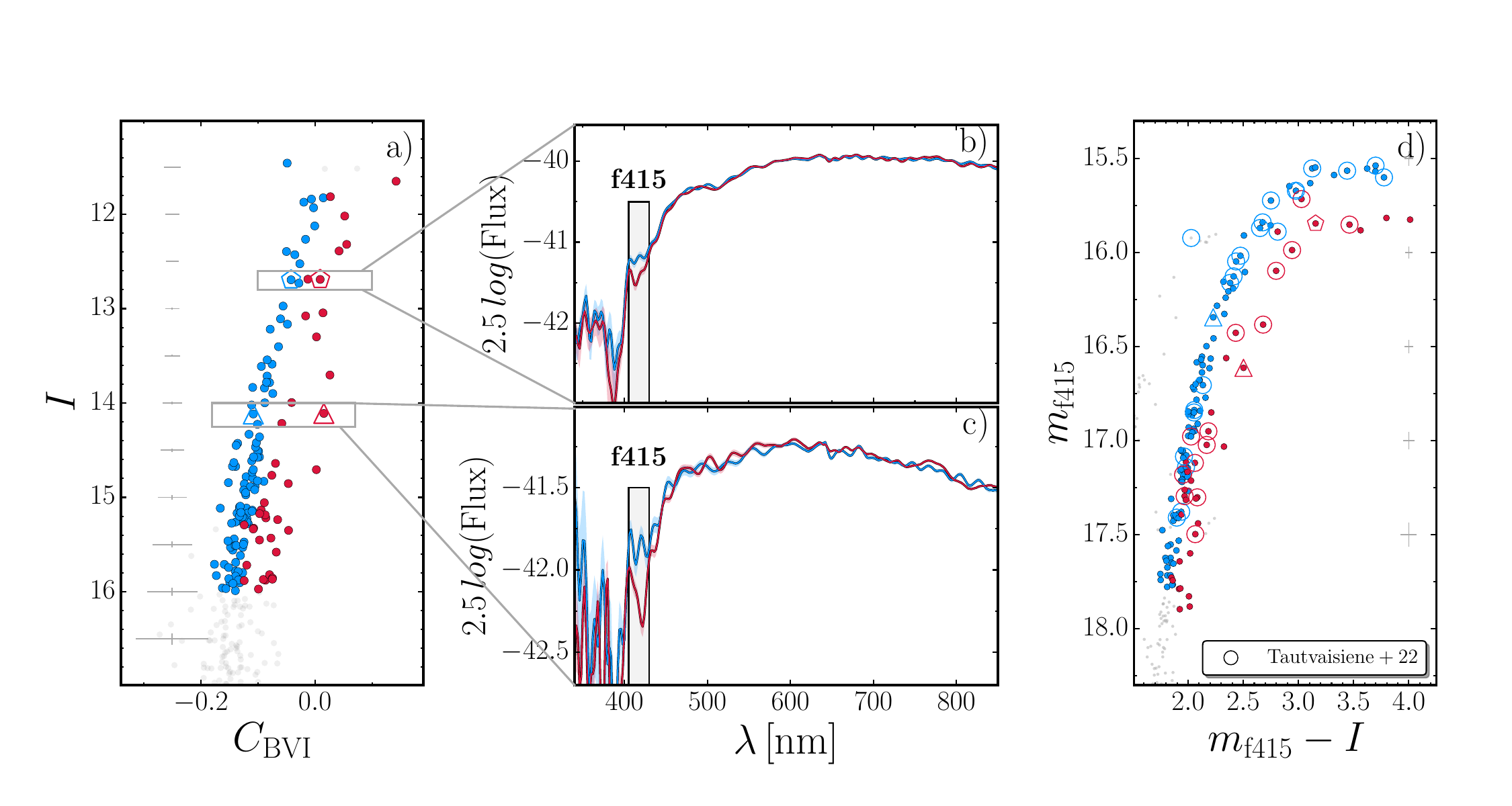}
    \caption{Effectiveness of the f415 photometric filter. \textit{Panel a).} $I$ vs. $C_{\rm BVI}$ CMD, as in Fig.~\ref{fig:popsel}a. Stars are colored azure and red according to the classification described in Sect.~\ref{sec:mp}, i.e. canonical and anomalous. Gray errorbars represent the average photometric uncertainties determined in different magnitude bins \textit{Panel b-c).} Externally calibrated Gaia XP spectra of two couples of reference canonical and anomalous stars, indicated with pentagons and triangles in panel a). The gray shaded rectangular regions indicate the $\rm f415^{25}$ filter, designed to identify canonical and anomalous stars. Fluxes are expresses in units of $\rm W m^{-2} nm^{-1}$. Azure and red shaded regions indicate flux uncertainties. \textit{Panel d).} $m_{\rm f415^{25}}$ vs. $m_{\rm f415^{25}}-I$, where $m_{\rm f415^{25}}$ is the magnitude computed by convolving externally calibrated spectra with the $\rm f415^{25}$ filter shown in panel a), as in Equation \ref{eqn:fnu}. Reference canonical and anomalous stars are enclosed by empty pentagons and triangles, while metal-poor and metal-rich stars from \citet{tautvaisiene2022} are marked with empty circles.} 
    \label{fig:spectra}
\end{figure*}

To convert the set of BP/RP coefficients into a single externally calibrated spectrum, we used the \texttt{GaiaXPy.calibrate} function. The resulting spectrum is expressed in units of $\rm W m^{-2} nm^{-1}$ and is dependent on the wavelength $\lambda$, measured in nanometers ($\rm nm$). Figures~\ref{fig:spectra}b-c depict the calibrated spectra with relative flux uncertainties for two pairs of reference canonical and anomalous stars, marked with pentagons and triangles, respectively, in Figure~\ref{fig:spectra}a.

A qualitative examination of the spectra in Figures~\ref{fig:spectra}b-c reveals that:
\textit{i)} The fluxes in the UV region ($\lambda \lesssim 400 \, \rm nm$) exhibit significant fluctuations.
\textit{ii)} Both sets of canonical and anomalous stars display notable flux differences in the spectral range between 400 and 430 $\rm nm$.
\textit{iii)} There are no discernible differences in the spectra for wavelengths $\lambda \gtrsim 450 \, \rm nm$.
For consistency, we checked that the same considerations are also true for different couples of canonical/anomalous stars selected randomly. 
Based on these findings, we have defined a box-shaped filter, denoted as $\rm f415^{25}$ and illustrated in Fig.~\ref{fig:spectra}b-c. The chosen name for the filter is based on indicating both its central wavelength and width. This filter, which encompasses the wavelength range between 405 and 430 $\rm nm$, is characterized by a rectangular transmission curve with a 100\% transmission efficiency across the specified wavelength range and zero outside of it.

To convolve the externally calibrated with the $\rm f415$ filter, we used the following equation:

\begin{equation}
    f_\nu = \frac{1}{c}\frac{\int S\cdot F \cdot \lambda d\lambda}{\int F/\lambda d\lambda} \qquad    
    \label{eqn:fnu}
\end{equation}
where $c$ is the speed of light in \AA$\rm/s$, $S$ is the stellar spectrum in $\rm erg s^{-1} cm^{-2}$\AA$^{-1}$, $F$ is the filter transmission curve and $\lambda$ represents the wavelength in units of \AA. We then computed the magnitude as 
$$m_{\rm f415} = -2.5\cdot log(f_\nu) - 48.6$$
% Figure~\ref{fig:spectra}d shows the resulting $m_{\rm f415}$ vs. $m_{\rm f415}-I$ CMD, with $I$ being the $I$ magnitude computed in the standardized JCK photometric systems, as discussed in Sect.~\ref{subsec:xp spectra}. Reference canonical and anomalous stars selected in panel a) are marked again with pentagons and squares, while metal-poor and metal-rich stars identified in \citet{tautvauisiene2022} are indicated by empty large azure and red circles, respectively.  \\
% Clearly, the ad-hoc designed $\rm f415$ photometric filter is highly effective in separating canonical and anomalous stars of bright RGB stars. Specifically, the color separation of canonical and anomalous stars ranges from 0.1 magnitudes in the faint RGB end, to nearly 0.4 in in mag for stars with $m_{\rm f415}\sim 16.X $. \\
% Furthermore, the UV spectral region is fundamental to identify multiple stellar populations with light-elements abundance variations, i.e. first and second populations as demonstrated by the high efficiency of the Johnson $U$ filter, and the HST $\rm F275W$ and $\rm F336W$ filters, in separating first- and second-population stars in GCs.\\
% Therefore, we attempted at selecting spectral features contained within 350 and 390 nm by defining a second ad-hoc photometric filter. However, the low quality of the spectra does not allow to properly compute synthethic magnitudes, hence preventing us from using this spectral region to investigate multiple populations with light-elements abundance variations. 

In Figure~\ref{fig:spectra}d, we present the $m_{\rm f415}$ vs. $m_{\rm f415}-I$ CMD, where $I$ represents the magnitude computed in the standardized JKC photometric systems, as discussed in Section~\ref{subsec:xp spectra}. The canonical and anomalous stars selected in panel a) are once again denoted by pentagons and triangles, respectively. Additionally, the metal-poor and metal-rich stars identified in \citet{tautvaisiene2022} are indicated by empty large azure and red circles, respectively. Average photometric uncertainties, determined by means of the SNR as discussed in Sec.~\ref{subsec:xp spectra}, are shown in Fig.~\ref{fig:spectra}a-d) as gray errorbars.

The ad-hoc designed $\rm f415$ photometric filter effectively separates bright RGB stars into canonical and anomalous groups. The color separation between canonical and anomalous stars varies from 0.1 magnitudes at the at the bottom of the RGB to nearly 0.4 magnitudes for stars with $m_{\rm f415^{25}}\sim 16$. 
To further demonstrate the effectiveness of the f415 filter, we show in App.~\ref{app:f415} the verticalized $m_{\rm f415^{25}}$ vs. $m_{\rm f415^{25}} - I$ CMD and the verticalized color distribution in different magnitude bins. We refer to App.~\ref{app:f415} for a detailed description and analysis of the results. 
%This demonstrates the effectiveness of the $\rm f415$ filter in distinguishing between these stellar populations.

\subsection{Tentative identification of multiple populations with light-element variations}
\label{sec:custom spectra 2}

Moreover, the UV spectral region plays a crucial role in identifying MPs with variations in light-element abundances, as supported by the effectiveness of the Johnson $U$ filter, as well as the HST $\rm F275W$ and $\rm F336W$ filters in separating 1P and 2P stars in Galactic GCs \citep{milone2012, piotto2015, milone2017}.

Therefore, we attempted to define a second ad-hoc photometric filter encompassing the 350-390 nm wavelength range, dubbed $\rm f360^{\rm 25}$. As in the case of the $\rm f415^{25}$ filter, the name of the filter indicates both the central wavelength and the width of the filter. However, the low quality and large fluctuations of the spectra in the UV region hinders the accurate computation of synthetic magnitudes, thereby preventing us from utilizing this spectral region for investigating abundance by using photometry. The detailed analysis of the performance of the filter is carried out in App.~\ref{app:f360}.

\section{Properties and internal dynamics of multiple populations}
\label{sec:properties}

In this section, we will analyze the morphology and the kinematics of the canonical and anomalous stellar populations identified in Section~\ref{subsec:xp spectra}. % to investigate the origin of multiple stellar populations in NCG\,1851.
In this context, the outer regions of the cluster are particularly important. %In multi-generation scenarios, second-population stars form after the collapse of polluted material in the cluster center,  
In most formation scenarios 2P stars form in the cluster center, leading to the formation of a more centrally concentrated new stellar generation.
Consequently, first-population stars are expected to dominate the cluster's outskirts. 
% This also applies to anomalous stars, which may form from heavy element-enriched material produced in supernovae. On the other hand, in single generation scenarios, MPs are expected to share similar dynamical and morphological properties.

In addition, %recent 
theoretical and $N$-body dynamical simulations, %such as those conducted by \citet{mastrobuono2013, henault2015, vesperini2018, vesperini2021, tiongco2019, sollima2021}, 
have shown that the present-day morphology and internal dynamics of multiple populations are strongly influenced by the initial configuration of the stellar population. In other words, the dynamical evolution of multiple populations in single- and multi-generational scenarios may follow different paths. For example, in multi-generational scenarios, the anisotropy profile of first-population stars should differ significantly from that of second-population stars in the outermost regions of the cluster %. For a more detailed discussion of the dynamical pazureictions, please refer to 
\citep{mastrobuono2013, henault2015, vesperini2018, vesperini2021, tiongco2019, sollima2021}.

Recently, \citet{cordoni2020a, cordoni2020b} used Gaia DR2 to investigate the kinematics of seven type-I and two Type-II Galactic GCs and found evidence of different dynamics and morphologies in at least four of the analyzed clusters. Here, we adopt a similar approach to investigate the internal dynamics of canonical and anomalous stars in NGC\,1851 up to a distance of approximately 45 arcmin from the cluster center.
%APM2. Uniformare il raggio massimo dappertutto.

\subsection{Morphology of multiple populations}
\label{subsec:radial}

\begin{figure*}
    \centering
    \includegraphics[width=0.99\textwidth, trim={0cm 0cm 1.8cm 0cm}, clip]{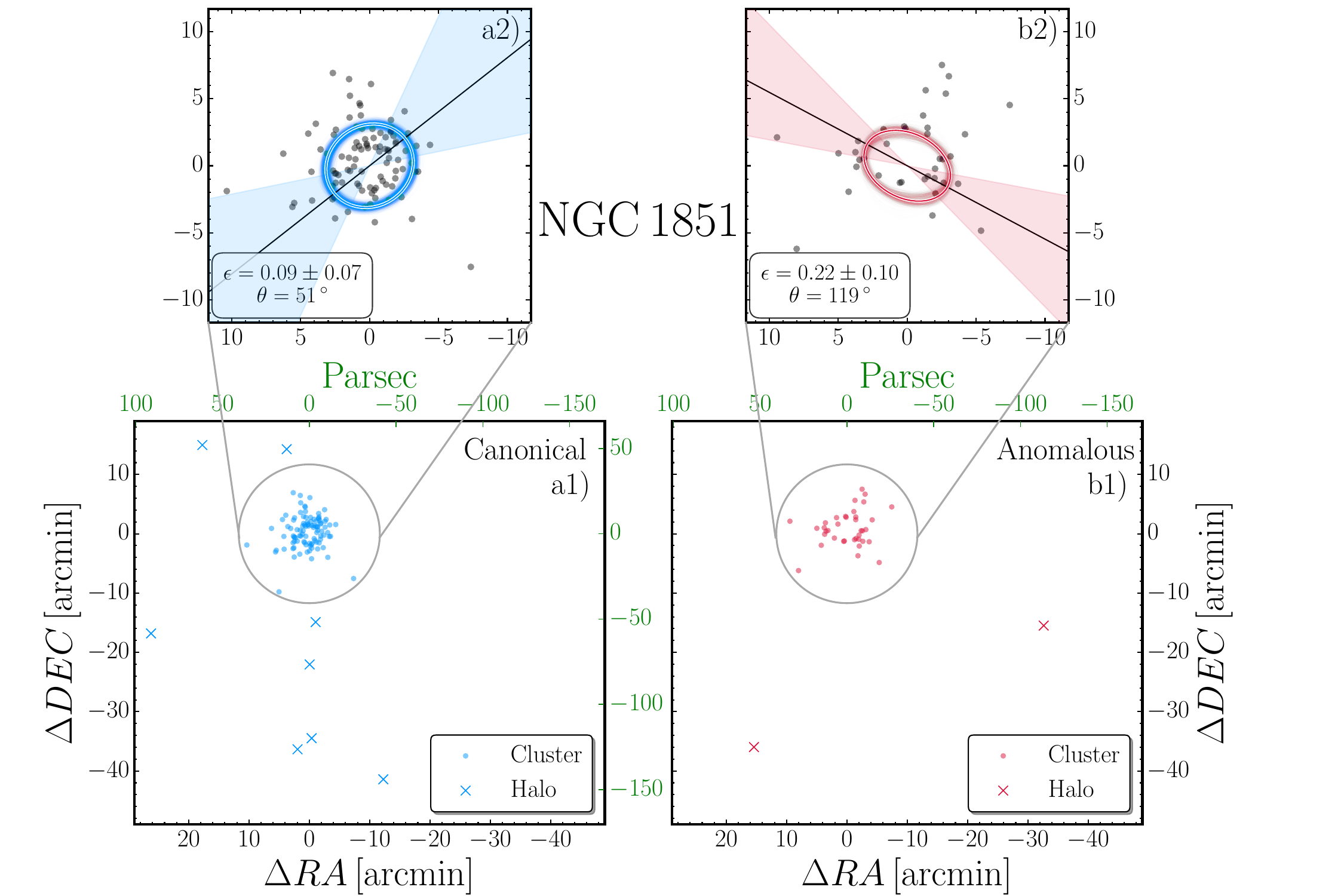}
    \includegraphics[width=0.8\textwidth, trim={1.5cm 0cm 0cm 0cm}, clip]{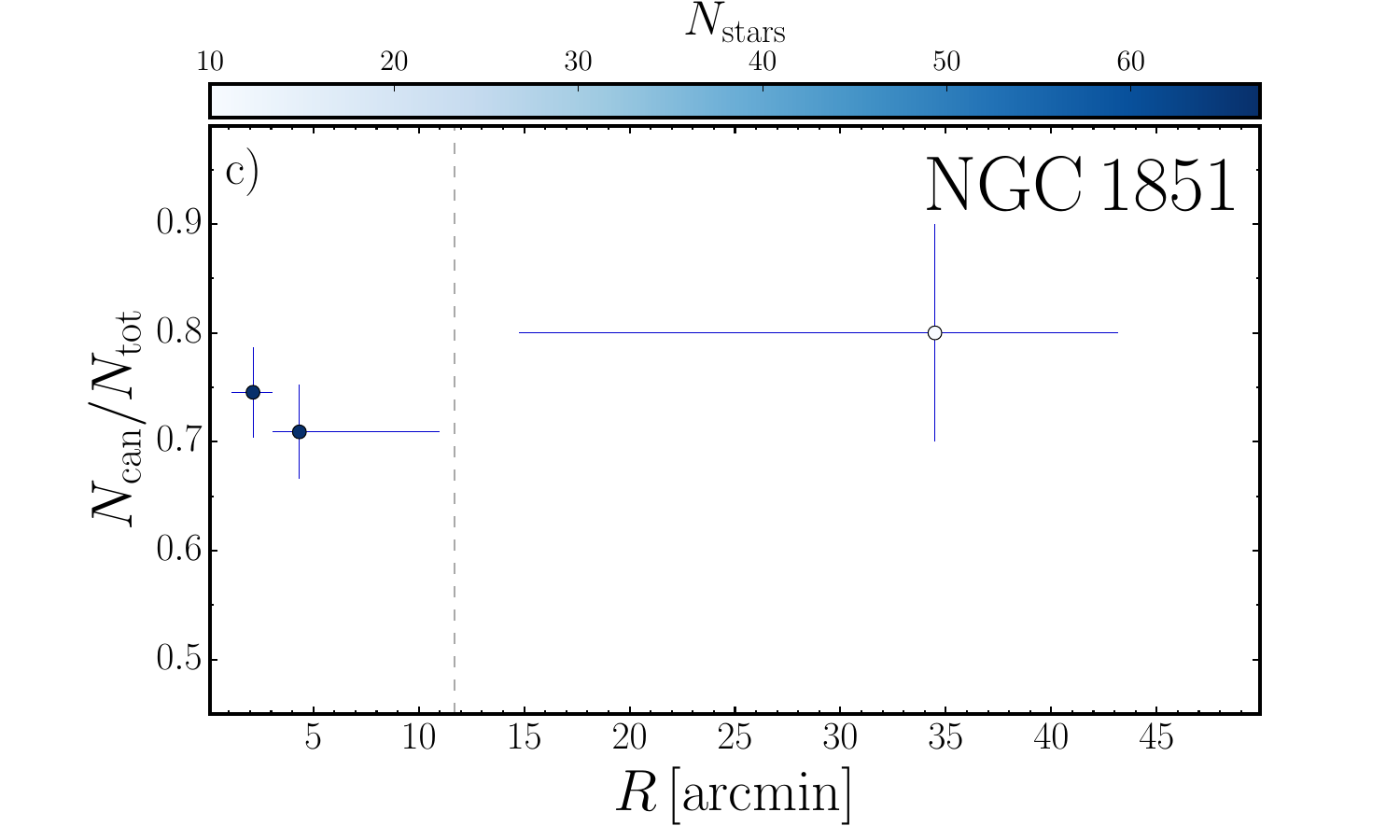}
    \caption{Morphology of multiple stellar populations. \textit{Panels a1-b1).} Spatial distribution of canonical and anomalous stars, shown respectively in azure and red colors. Stars within the tidal radius are marked as filled dots, while extra-tidal stars are represented with crosses. The 11.7 arcmin tidal radius is indicated by the gray circles. Physical units, i.e. parsecs, are displayed on the top and right axis in green. \textit{Panels a2-b2).} Zoom of the cluster field, with the best-fit ellipses shown as solid thick azure and red lines, and the position angle indicated by the solid black lines. Uncertainties are represented by the shaded regions. \textit{Panel c).} Radial distributions of canonical stars. The color is representative of the number of stars contained within each bin, while the gray dashed line indicates the tidal radius.}
    \label{fig: rad dist}
\end{figure*}

To analyze the spatial distribution of canonical and anomalous stars we fitted ellipses to the distribution of stars located within the tidal radius.
Indeed, as the tidal tail of NGC\,1851, mostly located in the southern quadrants \citep{kuzma2018}, exhibit spatial asymmetry (see e.g. Fig.~\ref{fig: rad dist}a1-b1), we excluded extra-tidal stars from the fitting procedure and fitted canonical and anomalous cluster stars within the tidal radius with an ellipse to determine their ellipticity ($\epsilon = 1-b/a$), position angle ($\theta$), and center. The uncertainties on the parameters were obtained by bootstrapping the results 1000 times. The best fit ellipses are shown in Fig.\ref{fig: rad dist}a2 and b2 with solid thick azure/red lines, and the bootstrapped fits with shaded thin lines. The black solid lines represent the inferred position angles, and the shaded azure/red regions indicate their uncertainties. %Due to the limited number of stars, we could not estimate the ellipticity profile, i.e. ellipticity as a function from radial distance from the cluster center, as in \citet[][see their Section 3]{cordoni2020a, cordoni2020b}. 

Our results suggest that the canonical and anomalous populations possibly have different morphologies, with the anomalous population being more elliptical (with $\epsilon = 0.22\pm0.10$) than the canonical population ($\epsilon=0.09\pm0.07$). However, the large relative uncertainties prevent us from drawing firm conclusions.
%APM2: questi numeri sono un po diversi da quelli di Figura 5. Aggiungere poi qui le barre d'errore.

% To study the radial distribution of multiple stellar populations in NGC\,1851, we divided the field of view in two annular regions containing approximately the same number of stars, and located within the tidal radius. Extra-tidal stars are instead contained in a separate bin.  For each bin, we computed the ratio between the number of canonical stars and the total number of stars. To test the influence of the binning choice, we repeated the analysis for different number of bins and varying the bin-overlap. We find that the results are consistent as long as a sufficient number of stars is contained within each bin, e.g. 20 stars.  The top panel of
% Fig.~\ref{fig:rad dist} shows the radial distribution of canonical stars in NGC\,1851, with each point color-coded according to the number of stars in contained within the bin. A visual inspection of Fig.~\ref{fig: rad dist}a reveals that, the population ratio inside the tidal radius is roughly constant within the  uncertainties, while it rises outside the tidal radius, reaching the value of $N_{\rm Can}/N_{\rm tot} = 0.85$. Specifically, we detect 11 canonical and 2 anomalous stars beyond the 11.7 arcmin tidal radius.  Hence, it seems that the halo of NGC\,1851 is dominated by stars belonging to the canonical populations. \\

To investigate the radial distribution of multiple stellar populations in NGC\,1851, i.e. the ratio between canonical and anomalous stars as a function of radial distance from the cluster center, we divided the region between 1.5 arcmin and the tidal radius into two circular annuli, each containing approximately the same number of stars, while all the extra-tidal stars are included in a unique annulus. We calculated the ratio of canonical stars over the total number of stars for each bin. We performed this analysis using different numbers of bins to evaluate the impact of binning choice, and we found that the results were consistent.

Figure~\ref{fig: rad dist}c displays the radial distribution of canonical stars in NGC\,1851. Each point is color-coded according to the number of stars contained in the bin. A visual inspection of Fig.\ref{fig: rad dist}c indicates that, at odds with \citet{zoccali2009}, the population ratio is relatively constant within the tidal radius, but increases beyond it, reaching a value of $N_{\rm Can}/N_{\rm tot} = 0.80\pm0.10$. Specifically, we detected eight canonical and two anomalous stars beyond the tidal radius of 11.7 arcmin. Therefore, it appears that the halo of NGC\,1851 is dominated by stars belonging to the canonical populations, in agreement with the finding of \citet{marino2014}. Nonetheless, we remind that given the low number of stars, the outer halo ratio is consistent with the average population ratio, i.e. $0.735\pm 0.036$. We suggest that the discrepancy with the results of \citet{zoccali2009} may be attributed to the asymmetric distribution of anomalous stars, which can introduce local gradients in the population ratio. 

%APM2. Qui bisogna aggiungere le barre d'errore e la referenza al lavoro di letteratura.
% Va anche specificata la regione analizzata, visto che noi in realta' non abbiamo stelle al centro.

\subsection{Internal dynamics of multiple stellar populations}
\label{subsec:dynamics}
% To investigate the internal dynamics of canonical and anomalous stars in the type-{\rm II} GC NGC\,1851, we follow a similar approach to the one described in \citet{vasiliev2018, bianchini2018, binachini2018b, cordoni2020a, cordoni2020b}. \\
% First we translate proper motions into sky-projected radial and tangential components, hereafter referazure to as $\mu_{\rm RAD}, \,\mu_{\rm TAN}$, converting uncertainties and covariances as well. Specifically, positive values of $\mu_{\rm RAD}$ indicate expansion, counter-clockwise rotation corresponds to positive $\mu_{\rm TAN}$. \\
% The radial and tangential dynamical profiles, i.e. mean motion and velocity dispersion profiles, are then computed minimizing the the log-likelihood function presented in \citet{bianchini2018}, updated to include the covariance term described in \citet{sollima2019}, as in \citet{bianchini2018b}. \\
% The canonical and anomalous populations are divided into respectively four and three annular bins with the same number of stars, and for each bin we minimize the likelihood by means of the \texttt{scipy} minimizer, using the Nelder-Mead method. Moreover, as for the radial distribution, canonical extra-tidal stars are treated in a separate bin.\footnote{Unfortunately, due to the small number of anomalous halo stars, we could not analyze them in a separate bin, as in the case of anomalous stars.}  \\
% Finally, we computed the anisotropy as $\beta=\sigma_{\rm TAN}/\sigma_{\rm RAD} -1$, so that tangentially anisotropic motion will translate into positive value of $\beta$. 
As mentioned in the previous paragraphs, the present-day internal dynamics of MPs provides an important vantage point on the formation of multiple populations and in turn of GCs. On the same note, the clusters' outskirts, which are characterized by longer relaxation times, are more likely to retain initial morphological and/or dynamical differences.  
In this work we adopt a similar approach to previous Gaia-based studies \citep[see e.g.][]{vasiliev2021, bianchini2018, bianchini2019, cordoni2020a, cordoni2020b} to investigate the internal dynamics of canonical and anomalous stars in NGC\,1851. Firstly, we convert the proper motions to sky-projected radial and tangential components, denoted as $\mu_{\rm RAD}$ and $\mu_{\rm TAN}$, respectively, and propagate their uncertainties and covariance. Positive $\mu_{\rm RAD}$ values indicate expansion, while positive $\mu_{\rm TAN}$ values indicate counter-clockwise rotation.
Next, we computed the radial and tangential dynamical profiles, i.e., mean motion and velocity dispersion profiles, by minimizing the log-likelihood function presented in \citet{bianchini2018}, with an updated covariance term from \citet{sollima2019}, as in \citet{bianchini2019}. The mean radial motion has been corrected for the expected apparent contraction or expansion due to the cluster's motion along the line-of-sight \citep[see e.g. Equation 2 and 3 of][]{bianchini2018}. We divided the canonical and anomalous populations into four and three annular bins with equal number of stars, respectively, and minimized the likelihood using the Nelder-Mead method with the \texttt{scipy} minimizer \citep{scipy}. Additionally, we treat extra-tidal canonical stars in a separate bin, while due to the limited number of anomalous halo stars (2 stars), we cannot analyze them separately.
Finally, we calculate the anisotropy parameter as $\beta=\sigma_{\rm TAN}/\sigma_{\rm RAD} -1$, where positive/negative values of $\beta$ correspond to tangentially/radially anisotropic motion.
\begin{figure}
    \centering
    \includegraphics[width=0.48\textwidth, trim={0cm 0cm 0cm 0cm}, clip]{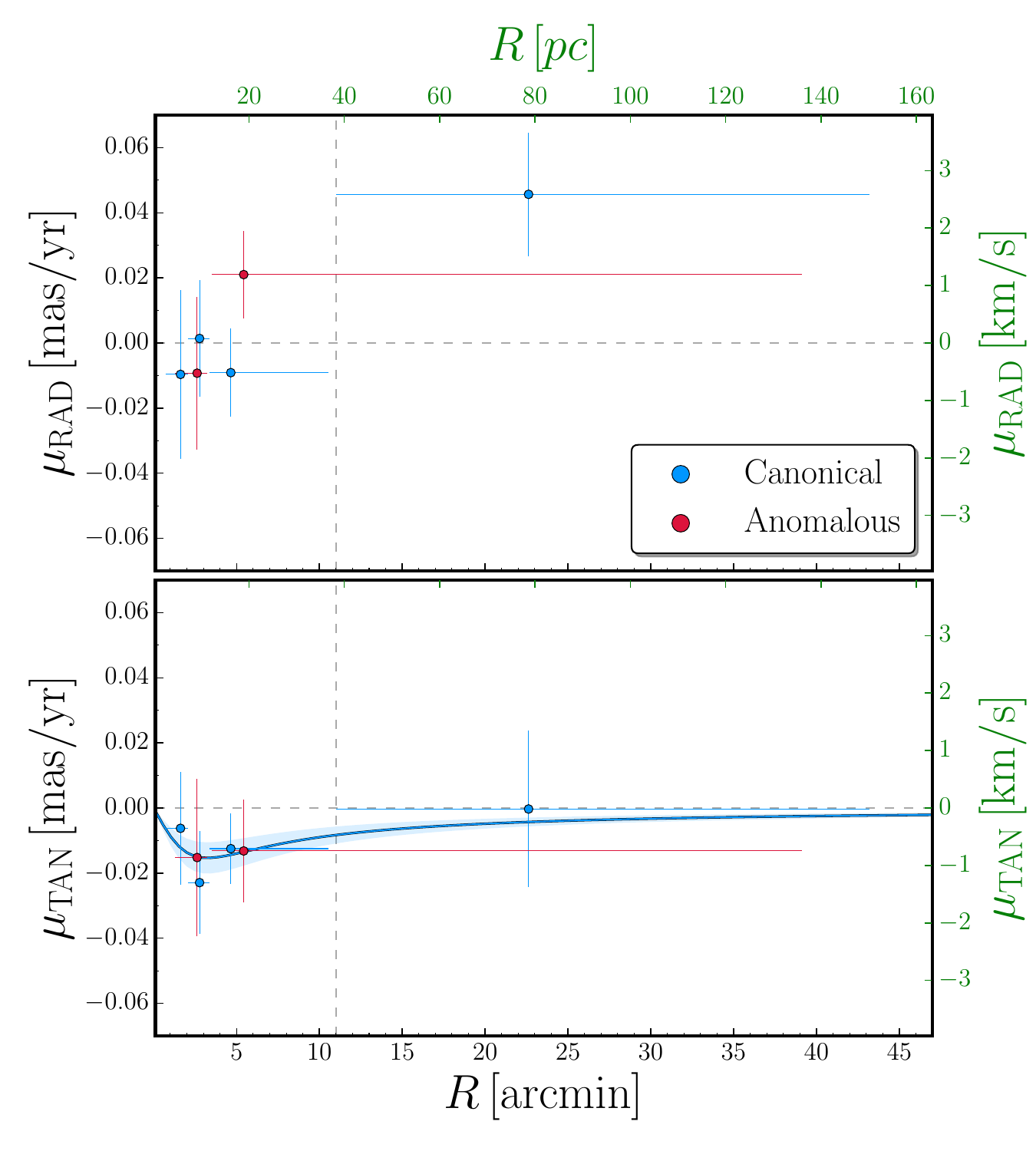}
    \caption{Mean motion of multiple stellar populations. Mean motion along the radial (top panel) and tangential (bottom panel) direction. Azure and red points indicate canonical and anomalous dynamical profile. Physical units (i.e. parsec and km/s) are displayed in the top and right axis in green.}
    \label{fig:dyn med}
\end{figure}

\begin{figure}
    \centering
    \includegraphics[width=0.48\textwidth, trim={0cm 0cm 0cm 0cm}, clip]{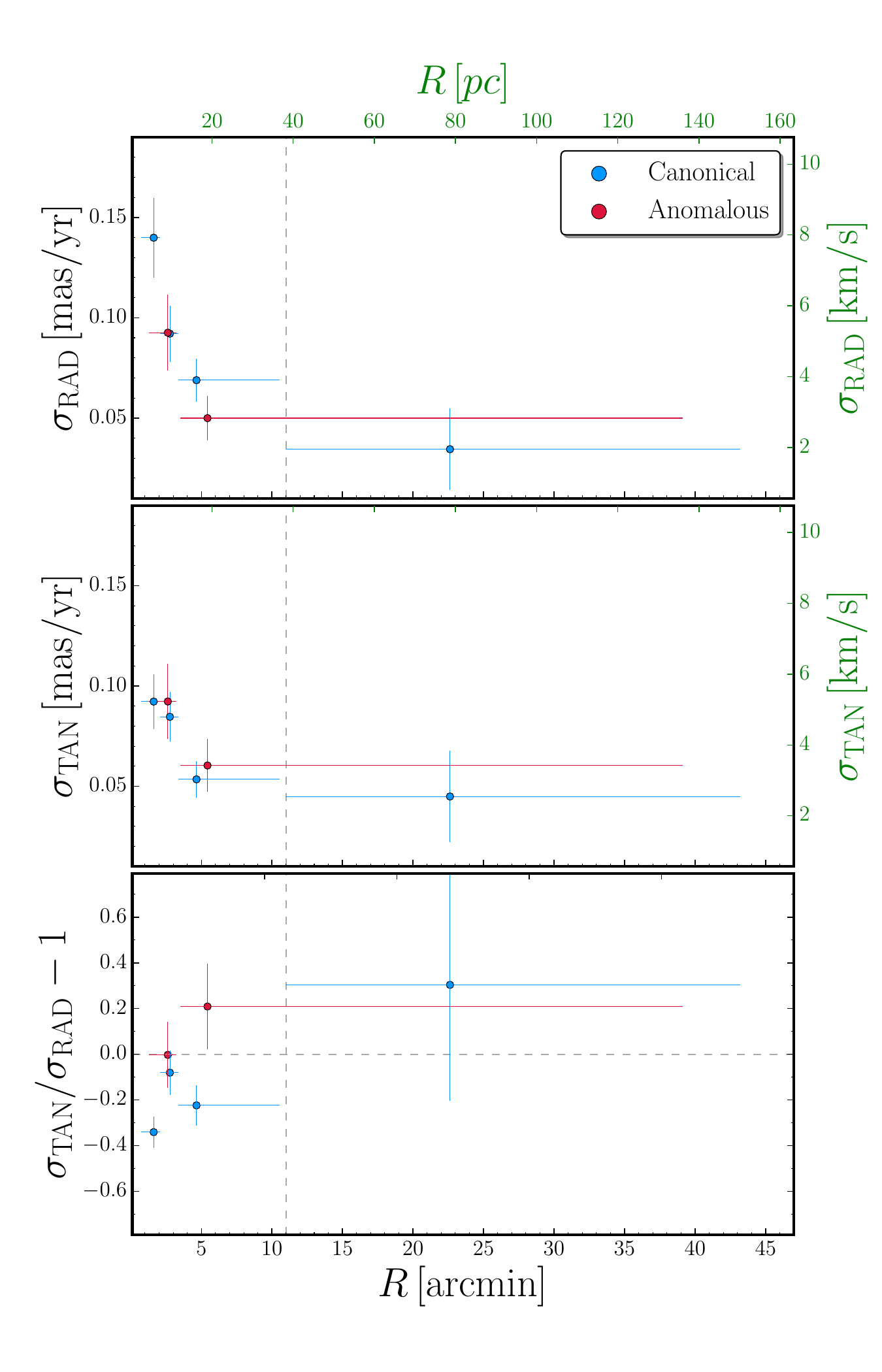}
    \caption{Velocity dispersion profile of multiple stellar populations. Dispersion profile along the radial (top panel) and tangential (middle panel) direction. Azure and red points indicate canonical and anomalous dynamical profile. Physical units (i.e. parsec and km/s) are displayed in the top and right axis in green. The anisotropy profile is shown in the bottom panel.}
    \label{fig:dyn dis}
\end{figure}

The uncertainties associated with the mean motion and velocity dispersion measurements were determined using the Markov Chain Monte Carlo algorithm \texttt{emcee} \citep{emcee}. The mean motion and dispersion profiles are presented in Figure~\ref{fig:dyn med} and Figure~\ref{fig:dyn dis}, respectively, for the canonical and anomalous stellar populations with azure and red colors. The distance from the cluster center is expressed in arcmin and parsecs, respectively on the bottom and top axes. Dynamical profiles are instead represented in mas/yr and km/s on the left and right axes. The green color is used to indicate parsecs and km/s, which have been converted adopting a distance of 11.95 kpc and a line-of-sight system velocity of 321.4 km/s \citep{baumgardt2019}.
Our analysis of NGC\,1851 reveals that canonical and anomalous stars share similar average motion along both the radial and tangential directions. Specifically, in the innermost regions of the cluster, both the canonical and anomalous stellar populations exhibit negligible radial motion. However, as we move outwards, there is a noticeable increase in positive radial motion, suggesting that stars in this region have a tendency to move on highly radial orbits, potentially escaping from the cluster.

In contrast, the mean motion along the tangential direction displays a specific %n interesting 
pattern. The central regions of NGC\,1851 exhibit negative tangential motions, indicating a clockwise rotation. On the other hand, extra-tidal stars show no significant rotation. This rotational pattern aligns with the expected rotation curve of GCs, \citep[see e.g.][]{mackey2013, kacharov2014, bianchini2018}, 
\begin{equation}
    \mu_{\rm TAN} = 2\frac{V_{\rm peak}}{R_{\rm peak}}\frac{R}{1+(R/R_{\rm peak})^2} 
    \label{eqn:rot}
\end{equation}
where the maximum rotational velocity is typically observed within a few arcminutes from the cluster center. The best-fit of the mean tangential motion of canonical stars is represented in Figure~\ref{fig:dyn med} as an azure solid line, while the shaded region indicates the corresponding uncertainties.

The velocity dispersion profiles in Fig.~\ref{fig:dyn dis} show a characteristic trend, decreasing as we move away from the cluster center in both the radial and tangential directions, which agrees with typical behavior observed in GCs. At the tidal radius, the dispersion reaches a value of around 0.04 mas/yr or 2 km/s.

Notably, when examining the canonical population, the innermost regions (within 3 arcmin) display a larger radial velocity dispersion compared to the tangential dispersion, while extra-tidal stars show a slightly larger tangential velocity dispersion. Moreover, although the anomalous population exhibits a slightly larger tangential dispersion compared to canonical stars, the dispersion profiles of both populations are consistent within their uncertainties.
%APM2

Additionally, it is noteworthy that the velocity dispersion in the outermost regions %does not drop to zero but rather 
reaches a plateau. This behavior, characterized by an increasing or flat dispersion profile in cluster's outskirts, deviates from the expected trend for LIMEPY/SPES models \citep{gieleszocchi2015, deboer2019, claydon2019}, %in agreement with the findings of 
and is similar to what is observed by 
\citet{zhen2023} %for three other Galactic GCs, namely 
in the GCs NGC\,1261, NGC\,1851, and NGC\,1904. A plausible interpretation for this observed dispersion profile is the tidal interaction with the Galaxy. We mention here that, a qualitative comparison with Figure 9 from \citet{zhen2023} reveals that we obtain a lower value of the velocity dispersion profile compared to theirs, e.g. 2 km/s against 5 km/s. On the other hand, the small number of stars, i.e. 8 canonical stars, prevent us from drawing firm conclusions.

The anisotropy profile of canonical and anomalous stars is presented in the bottom panel of Figure~\ref{fig:dyn dis}. In the case of the canonical population, the motion initially exhibits a radially anisotropic behavior in the innermost regions, at approximately 2-3 arcmin from the cluster center, while, as we move towards the cluster halo, the anisotropy possibly transitions to tangential. On the other hand, the anomalous population displays an isotropic motion at a distance of 3-4 arcmin from the cluster center, and exhibits hints of tangential anisotropy in the outer regions.

%These findings are in agreement with the results reported by \citet{libralato2023ApJ}, who analyzed the internal dynamics of MPs using HST proper motions for 57 GCs. They observed that 2P and anomalous stars exhibit isotropic motions at the cluster center but become radially anisotropic at approximately half-light radius, which corresponds to around 0.51 arcmin for NGC\,1851. On the other hand, 1P stars is consistent with being isotropic across their observed field of view, spanning approximately 1.5 half-light radii. We remind the reader that, while \citet{libralato2023ApJ} investigates the regions within 1.5 half-light radii, in this work we only consider stars that lies outside $\sim 4$ half-light radii. \\
%APM2?
These findings suggest that the stellar populations of NGC\,1851 follow a typical behavior observed in other GCs \citep[e.g.][]{bellini2015, milone2018, cordoni2020a}. 
Similar conclusions are provided by \citet{libralato2023ApJ} who analyzed the proper motions of 1P, 2P and anomalous stars in the inner $\sim$2.7$\times$2.7 arcmin regions of 56 GCs and concluded that, when combining together the results from all dynamically old clusters i.e. $Age/t_{\rm rh }>10$\footnote{In the case of NGC\,1851 $Age/t_{\rm rh}\sim 14$}, 2P stars exhibit isotropic motions at the cluster center but become radially anisotropic at approximately half-light radius. On the other hand, 1P and anomalous stars are consistent with being isotropic across their observed field of view \citep[see e.g. the top-row panels in Fig.~3 of][]{libralato2023ApJ}.

% Furthermore, this behavior aligns with the findings of \citet{bianchini2019} for NGC\,3201, where the outermost regions are characterized by radially anisotropic motions. In the cluster outskirts, both the canonical and anomalous populations exhibit radially anisotropic motions. In contrast, within the innermost regions, the anomalous stars display isotropic motions, while the canonical stars exhibit tangential anisotropy.

To evaluate the impact of binning on the results, the dynamical profiles of the canonical stars were recomputed for different number of bins. Consistent results were obtained across different binning choices. Overall, the mean motions of the canonical and anomalous stars are similar, with hints of possible differences concerning the tangential dispersion profiles.

% Finally, the anisotropy profile of canonical and anomalous stars is depicted in the bottom panel of Figure~\ref{fig:dyn dis}. Concerning the canonical population, the motion starts as radially anisotropic in the innermost regions (e.g. within 2 arcmin) and grows toward tangential anisotropy in the cluster halo. A similar behavior is exhibited by the anomalous population, which is consistent with being isotropic at 3-4 arcmin from the cluster center, and radially anisotropic in the outer regions. Such results are consistent with the findings of \citet{libralato2023ApJ}. Specifically, analyzing the internal dynamics of MPs with HST proper motion for 57 GCs, Libralato and collaborators found that 2P stars exhibit isotropic motions at the center, but become radially anisotropic at half-light radius, i.e. 0.XX for NGC\,1851. On the other hand, 1P stars are consistent with being isotropic over their field of view, i.e. approximately 1.5 half-light radii. 

\begin{figure}
    \centering
    \includegraphics[width=0.45\textwidth, trim={0cm 2cm 0cm 2cm}, clip]{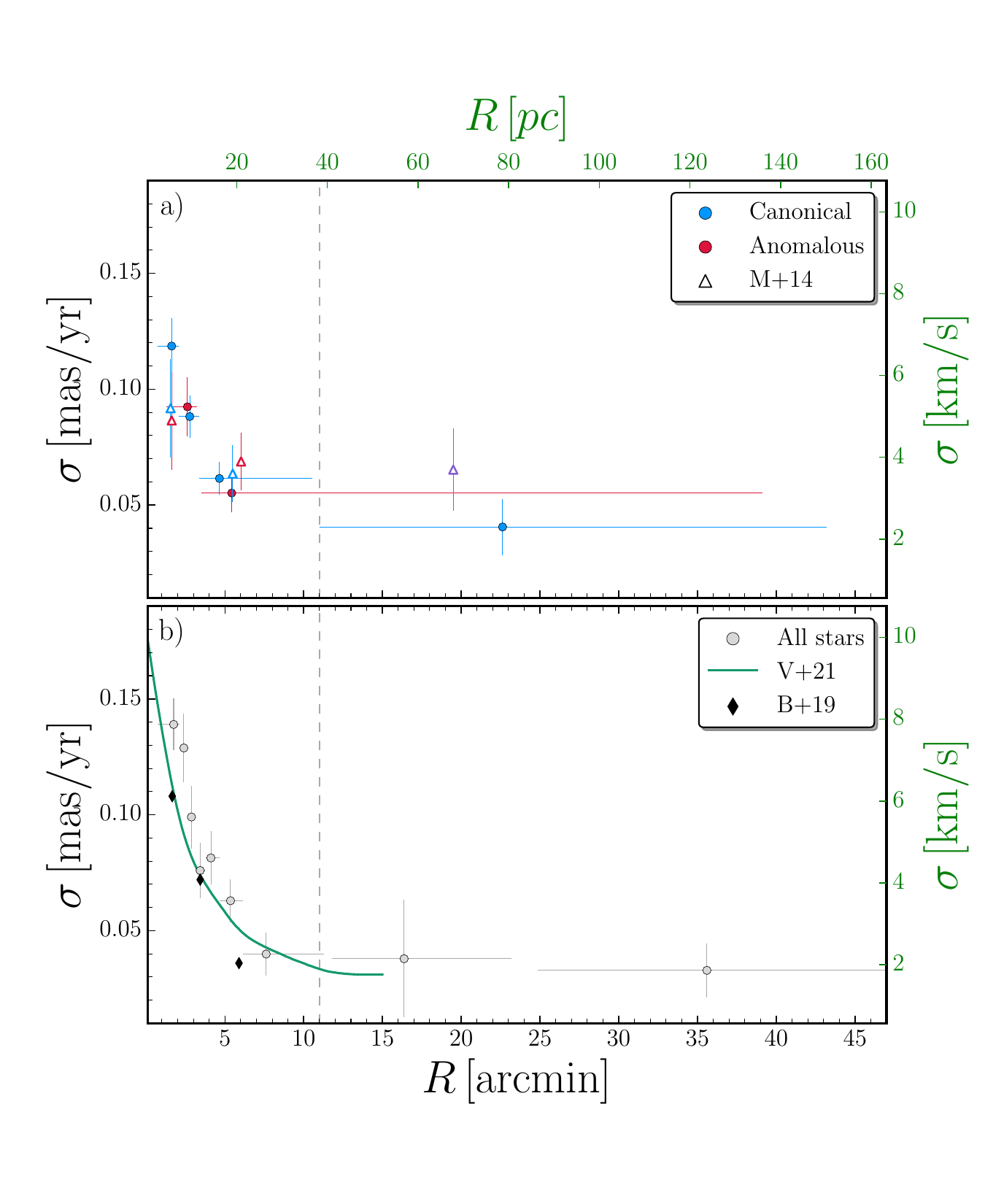}
    \caption{Comparison with literature 1D velocity dispersion profiles. \textit{Panel a).} Comparison with literature velocity dispersion profiles, determined from line-of-sight velocity measurements, of canonical and anomalous stars from \citep{marino2014}, respectively shown with azure and red empty triangles. Extra-tidal stars dispersion in indicated by the purple empty triangle. \textit{Panel b).} 1D velocity dispersion profile from proper motions from \citet{baumgardt2019, vasiliev2021}, represented by the black diamonds and green line, respectively. Dynamical profiles have been determined from all cluster stars. Grey circles indicate the 1D velocity dispersion computed in this work analyzing all cluster stars, regardless of their canonical/anomalous classification.}
    \label{fig:literature}
\end{figure}

In Figure~\ref{fig:literature}, we present a comparison of our 1D velocity dispersion profiles with previous literature studies. Additionally, we show the overall 1D dispersion profile, computed for all stars, regardless of their canonical/anomalous identification, as gray markers in Figure~\ref{fig:literature}. Such profile can be directly compared with the profile obtained by \citet{vasiliev2021}, denoted by the green line. Moreover, Figure~\ref{fig:literature} also shows the dispersion profile of canonical and anomalous stars determined in \citet{marino2014} from  line-of-sight velocities,  and the overall dispersion profile of \citet{baumgardt2019}, obtained combining Gaia DR2 proper motions with line-of-sight velocities. A visual examination of Figure~\ref{fig:literature} suggests a good overall agreement between our 1D velocity dispersion profiles and those reported in the literature.

% Although the canonical and anomalous populations exhibit similar velocity dispersion profiles, differences are detected in their anisotropy profiles. Specifically, the canonical population displays mild tangential anisotropy in the central regions, while the extra-tidal region exhibits radially anisotropic motions. However, the uncertainties associated with the extra-tidal regions limit the ability to draw definitive conclusions. 
%\section{Discussion and conclusion}
%APM e' piu' che altro un riassunto.
\section{Summary and conclusion}
\label{sec:conclusion}

% We exploited Gaia DR3 to identify canonical and anomalous stars in NGC\,1851 and investigate their morphological and dynamical properties. Specifically, using the python package \texttt{GaiaXPy} we converted Gaia DR3 low-resolution XP spectra into synthethic photometry in the JKC standardized photometric system \citep[see e.g.][for a detailed description of the photometri system]{montegriffo2022b}, i.e. $UBVI$ photometry. 

% By following the procedure described in \citet{milone2018, cordoni2018, cordoni2020} and adopting the recipes of \citet{riello2018}, we separated NGC\,1851 cluster members from contamining interlopers. Finally, we exploit the $C_{\rm BVI}=(B-V)-(V-I)$ photometric index, to identify canonical and anomalous stars, i.e. metal-poor and metal-rich stars,  among bright RGB stars. To demonstrate the effectiveness of the $C_{\rm BVI}$ index in mapping the heavy-elements content of RGB stars, we cross-matched our identification with the one performed in \citet{tautvaisiene2022}, which is based on spectroscopic measurements. We find that our photometric classification is in agreement with the spectroscopic one. \\

We investigated the morphological and dynamical properties of multiple populations in NGC\,1851 by using Gaia Data Release 3 \citep{gaiadr3}. To accomplish this, we made use the python package \texttt{GaiaXPy} to convert Gaia DR3 low-resolution XP spectra into synthetic photometry in the JKC standardized photometric system  %,  specifically following the recipe described in 
 \citep{montegriffo2022a, montegriffo2022b} and generating $UBVI$ photometry.

Specifically, we derived the $C_{\rm BVI}=(B-V)-(V-I)$ index, which is a powerful tool to identify stellar populations with different metallicities among the bright RGB stars of GCs \citep{marino2015}.
Moreover, we %exploit Gaia DR3 low-resolution XP spectra to 
define a new photometric filter, here dubbed f415, %(see Figure~\ref{fig:spectra}b and c), 
which isolates a distinctive feature of canonical and anomalous stars, located approximately at 400\,nm and maximizes the separation between the canonical and anomalous stars (i.e. $s$-poor and $s$-rich stars). %Notably, the photometry generated with the f415 filter correctly distinguish canonical and anomalous stars (see Figure~\ref{fig:spectra}d).

%Following the methodology outlined in \citet{milone2018, cordoni2020a, cordoni2020b} and adopting the recipes from \citet{riello2018}, we implemented a procedure to distinguish NGC\,1851 cluster members from contaminating field stars. 
%Furthermore, we exploited the $C_{\rm BVI}=(B-V)-(V-I)$ photometric index to identify canonical and anomalous stars, i.e. metal-poor and metal-rich stars, respectively, among bright RGB stars in NGC\,1851. %To validate the effectiveness of the $C_{\rm BVI}$ index in identifying canonical and anomalous stars, we performed a cross-match with the classification of \citet{tautvaisiene2022} based on spectroscopic measurements. Remarkably, our photometric classification aligns well with the spectroscopic classification. 

%Moreover, we exploit Gaia DR3 low-resolution XP spectra to define an ad-hoc photometric filter, here dubbed f415 (see Figure~\ref{fig:spectra}b and c), which isolate a distinctive feature of canonical and anomalous stars, located approximately at 400\,nm. Notably, the photometry generated with the f415 filter correctly distinguish canonical and anomalous stars (see Figure~\ref{fig:spectra}d).  

%By using the $C_{\rm BVI}$ photometric index, we successfully identified and classified canonical and anomalous stars in NGC\,1851, demonstrating the reliability and agreement of our photometric approach with spectroscopic measurements. 
%Moreover, we adopted a procedure similar to that described in \citet{cordoni2020a, cordoni2020b} to investigate the radial and spatial distribution of these populations. Our findings are as follows:
The main results on the morphology of multiple populations and on their radial and spatial distributions include:
\begin{itemize}

    \item The overall fraction of canonical stars contained within the tidal radius is $0.730 \pm 0.038$, and is consistent with results from previous studies \citep[e.g.][]{milone2009}.
    %APM. 
    %The overall ratio of canonical to anomalous stars within the tidal radius is consistent with previous studies based on HST photometry, i.e. $N_{\rm Can}/N_{\rm tot} = 0.65 \pm 0.XX$ \citep{milone2009}. We find a ratio of $N_{\rm Can}/N_{\rm Tot}=0.730 \pm 0.038$.

    \item 
    We confirm the presence of an halo of  cluster members that surrounds NGC\,1851 and extends beyond the tidal radius up to more than 45 arcmin.  The halo is predominantly composed of canonical stars, which comprise the $80 \pm 10 \%$ of the total number of stars. This finding is in line with the conclusions of \citet{marino2014}, who, based on an analysis of spectra from the Very-Large Telescope for seven extra-tidal stars, also reported that the NGC\,1851 halo is primarily composed of the canonical population.
    % ho tolto questa parte
    % This result is consistent with the conclusion by \citet{marino2014}, who analyzed Very-Large Telescope spectra of seven stars and find that all of them are consistent with belonging to the canonical population.
    %The halo of NGC\,1851, extending well beyond the 11.7 arcmin tidal radius in the south direction, is predominantly composed of canonical stars, with a ratio of $N_{\rm Can}/N_{\rm Tot}=0.80 \pm 0.10$. This confirms the earlier conclusions of \citet{marino2014}, who used high-resolution spectra to determine that all extra-tidal stars in their sample were consistent with belonging to the canonical population.

    \item The canonical and anomalous population exhibit a flat radial distribution in the entire field of view, in agreement with \citet{milone2009}.
    While the canonical population has a nearly circular distribution, the anomalous stars show elliptical shapes.
    %APM: i risultati sono sempre consistenti ad un sigma ed i punti sono solo tre. Per cui non parlerei di possibili gradiendi.
    %The radial distribution of canonical and anomalous stars exhibits a flat trend within the tidal radius, with an increasing trend towards the cluster outskirts. It is worth noting that potential spurious gradients may arise due to the analysis of different mass ranges for canonical and anomalous stars, as discussed in Section~\ref{subsec:radial}. %GC decidere se lasciiare ed aggiungere qualcosa oppure togliere
   % \item To investigate the on-sky morphology of canonical and anomalous stars, we fitted ellipses to the distributions of the two populations. Our analysis suggests that the anomalous population is more elliptical, while the canonical population is nearly circular. However, the large uncertainties associated with the best-fit ellipses prevent us from drawing definitive conclusions.
\end{itemize}

We analyzed the proper motions of the canonical and anomalous populations over the entire field of view.
In particular, we expand the analysis of the kinematics of multiple stellar populations beyond the tidal radius for the first time.
This is crucial % investigation of internal dynamics in multiple stellar populations is 
information for understanding the origin of GCs and their multiple stellar populations. %, as discussed in Section~\ref{subsec:dynamics}. 
Indeed, while dynamical differences in the inner regions may be erased by relaxation processes, the outskirts of the cluster, characterized by longer relaxation timescales, are more likely to retain evidence of distinct dynamical evolution. 
 
%Additionally, we utilize highly accurate HST proper motion measurements from \citet{libralato2022} to probe the innermost regions, which are not accessible with ground-based and Gaia data. Thus, we investigate the internal dynamics of canonical and anomalous stars, spanning from the inner arcseconds to nearly 50 arcminutes from the cluster center. 
Our findings can be summarized as follows:

\begin{itemize}
    \item Canonical and anomalous stars exhibit similar mean motions in the radial and tangential directions. Specifically, both populations display positive radial motions in the outskirts, indicating a tendency for stars to escape the cluster on radial orbits. The tangential velocity profile, which serves as a proxy for on-sky rotation, shows a peak at approximately 3 arcmin (i.e. $\sim 10$ parsec), followed by a decline to zero outside the tidal radius. Such pattern is consistent with the expected rotation of GCs.

    \item The canonical population appears to have a slightly lower tangential velocity dispersion compared to the anomalous population, while no significant differences are observed in the radial velocity dispersion profile.

    \item The velocity dispersion profiles exhibit a flat/increasing trend in the outskirt regions. This results, which is similar to findings of \citet{zhen2023} for the GCs NGC\,1261, NGC\,1851, and NGC\,1904, is in disagreement with the expectations from LIMEPY/SPES models \citep{gieleszocchi2015, deboer2019, claydon2019}.
    %This contradicts the expectations from LIMEPY/SPES models \citep{gieleszocchi2015, deboer2019, claydon2019}, while it agrees with the findings of \citet{zhen2023} for three Galactic GCs.

    \item In the outskirts of the cluster, both canonical and anomalous populations display some hints of tangentially anisotropic motion. %However, their anisotropy profile seems to diverge in the central regions, where 
    In contrast, in the innermost regions, 
    the canonical population shows radial anisotropy, while anomalous stars exhibit isotropic motion. %These distinctions persist regardless of the adopted binning choice. 
    %Nonetheless, the small number of stars and relatively high uncertainties prevent us from drawing firm conclusions.
\end{itemize}

% In summary, our analysis sheds light on the internal dynamics of canonical and anomalous stars in NGC\,1851, revealing distinct characteristics in different regions of the cluster. These findings contribute to our understanding of the dynamical processes and evolution of multiple stellar populations within globular clusters.

In general, our study demonstrates that Gaia DR3 low-resolution XP spectra, together with Gaia DR3 astrometry and proper motions are powerful tools to investigate multiple populations in GCs.
We provided a comprehensive analysis of the canonical and anomalous populations in the type-II GC NGC\,1851 and its halo. %. Moreover, we provide valuable 
%We provide insights into 
The distinct behaviors in the morphology and in the internal dynamics of canonical and anomalous stars, %highlighting their distinct behaviors 
have been followed
in different regions of the cluster. These findings enhance our understanding of the dynamical processes and evolution of multiple stellar populations within globular clusters.

\section*{Acknowledgements}
We thank the anonymous referee for his/her comments which improved the quality of the manuscript. AFM, GC and APM acknowledge the support received from INAF Research GTO-Grant Normal RSN2-1.05.12.05.10 - Understanding the formation of globular clusters with their multiple stellar generations (ref. Anna F. Marino) of the "Bando INAF per il Finanziamento della Ricerca Fondamentale 2022". This work has received funding from the European Union’s Horizon 2020 research and innovation programme under the MarieSklodowska-Curie Grant Agreement No. 101034319 and from the European Union – NextGenerationEU, beneficiary: Ziliotto.
This work has made use of data from the European Space Agency (ESA) mission
{\it Gaia} (\url{https://www.cosmos.esa.int/gaia}), processed by the {\it Gaia}
Data Processing and Analysis Consortium (DPAC,
\url{https://www.cosmos.esa.int/web/gaia/dpac/consortium}). Funding for the DPAC
has been provided by national institutions, in particular the institutions
participating in the {\it Gaia} Multilateral Agreement. This job has made use of the Python package GaiaXPy, developed and maintained by members of the Gaia Data Processing and Analysis Consortium (DPAC), and in particular, Coordination Unit 5 (CU5), and the Data Processing Centre located at the Institute of Astronomy, Cambridge, UK (DPCI). SJ acknowledges support from the NRF of Korea (2022R1A2C3002992, 2022R1A6A1A03053472)

% %%%%%%%%%%%%%%%%%%%%%%%%%%%%%%%%%%%%%%%%%%%%%%%%%%
% \section*{Data Availability}
The data underlying this article will be shared on reasonable request to the corresponding author.

\bibliographystyle{aa}
\bibliography{ms} % if your bibtex file is called example.bib

\begin{appendix}

\section{Performance of the $\rm f360^{25}$ and $\rm f415^{25}$ ad-hoc photometric filters }
In the following, we will review in more details the performance and effectiveness of the two newly defined photometric filters, namely the $\rm f360^{25}$ and $\rm f415^{25}$. We remind here that, while the first is defined to reflect variations in light-elements abundance, i.e. to separate 1P from 2P stars, the latter aims at effectively distinguishing canonical and anomalous stars, i.e. stars with different heavy-elements content.

\subsection{Effectiveness of the $\rm f415^{25}$ photometric filter}
\label{app:f415}

\begin{figure*}
    \centering
    \includegraphics[width=0.8\textwidth, trim={0cm 0cm 0cm 0cm}, clip]{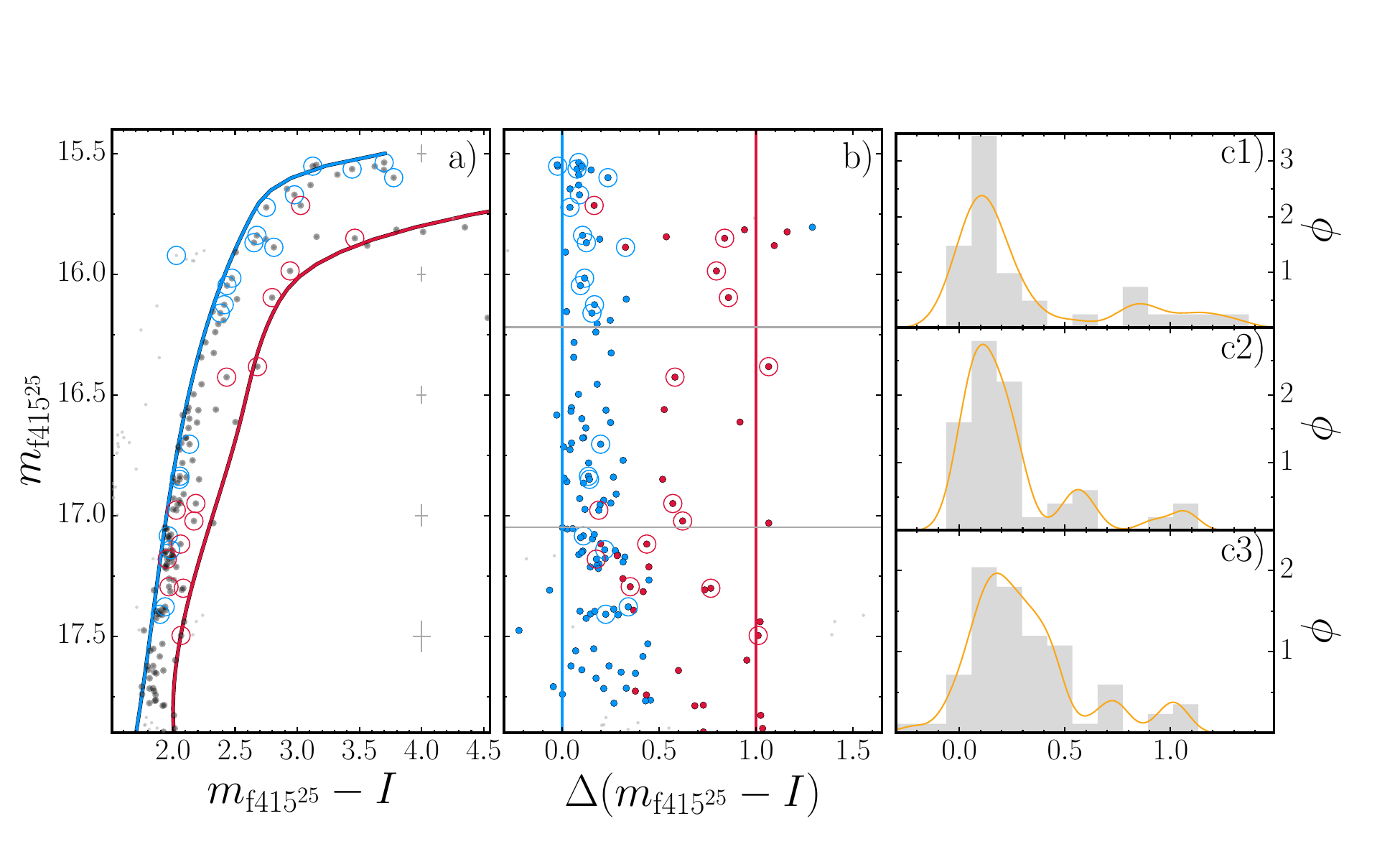}
    \caption{Effectiveness of the $\rm f415^{25}$ photometric filter. \textit{Panel a).} $m_{\rm f415^{25}}$ vs. $m_{\rm f415^{25}}-I$ CMD. Canonical and anomalous stars, selected by means of the $BVI$ index, are colored azure and red respectively. The azure and red fiducial lines mark the RGB boundaries, defined as the $4^{\rm th}$ and $96^{\rm th}$ percentile of the pseudo-color distribution. Average color and magnitude uncertainties are shown with the gray errorbars on the right of the CMD. Stars in common with \citet{tautvaisiene2022} are marked with empty large circles.  \textit{Panel b).} Verticalized CMD of panel a). \textit{Panels c1-c3).} Verticalized color distribution of stars in the three different magnitude ranges highlighted in panel b). The orange lines represent the Gaussian KDE.}
    \label{fig:f415}
\end{figure*}

Fig.\ref{fig:f415} presents a more detailed analysis of the effectiveness of the $\rm f415^{25}$ filter in distinguishing between canonical and anomalous stars, i.e. stars with different heavy-element content. To evaluate the performance of the $\rm f415^{25}$ filter, we adopt the same approach used for identifying MPs in the $I$ vs. $C_{\rm BVI}$ pseudo-CMD, shown in Fig.\ref{fig:popsel}. Specifically, we define the blue and red RGB boundaries as the $4^{\rm th}$ and $96^{\rm th}$ percentiles of the $m_{\rm f415^{25}}-I$ distribution (Fig.\ref{fig:f415}a), and calculate the verticalized $\Delta(m_{\rm f415^{25}}-I)$ (Fig.~\ref{fig:f415}b) as discussed in Sec.\ref{sec:mp}. Finally, we analyze the distribution of the verticalized color in three different magnitude bins indicated by the gray horizontal lines in Fig.~\ref{fig:f415}b.

Fig.\ref{fig:f415}c1-c3) illustrate the distribution of the verticalized $\Delta(m_{\rm f415^{25}}-I)$ color in the three magnitude bins. Remarkably, the $\Delta(m_{\rm f415^{25}}-I)$ distribution exhibits well separated peaks in all three magnitude bins, with the faintest bin displaying a certain degree of overlap. \\
Fig.\ref{fig:f415} suggests that the synthetic magnitude in the $\rm f415^{25}$ filter offers a clear identification of canonical and anomalous stars among bright RGB stars. However, for fainter RGB stars, the combination of increasing photometric uncertainties (as shown by the gray error bars in Fig.\ref{fig:f415}a), larger flux fluctuations, and a narrower CMD diminishes the clarity of separation, resulting in some overlap between canonical and anomalous stars. Furthermore, the less pronounced separation among faint-RGB stars is partly attributed to the intrinsic nature of canonical and anomalous stars. Firstly, the distribution of $\rm C+N+O$ is continuous rather than discrete \citep[see e.g.,][]{yong2015,tautvaisiene2022}, contributing to the reduced differentiation. Secondly, the separation also depends on stellar temperature, resulting in decreased distinction for colder faint-RGB stars. Similarly, this effect is evident in the $C_{\rm BVI}$ index, which more effectively identifies canonical and anomalous stars among bright RGB stars.

%Moreover, Fig.\ref{fig:f415} reveals that the synthetic magnitude in the $\rm f415^{25}$ filter allows for a clear identification of canonical and anomalous stars among bright RGB stars. However, for fainter RGB stars, the combination of increasing photometric uncertainties (as indicated by the gray error bars in Fig.~\ref{fig:f415}a), larger flux fluctuations, and a narrower CMD make the separation less evident, contribute to generate overlap between canonical and anomalous stars. Furthermore, we suggest that the less clear separation in the faint-RGB is partly intrinsic to the canonical/anomalous separation. Specifically, firstly, the $\rm C+N+O$ distribution is not discrete but rather continuous \citep[see e.g.][]{yong2015}, and then, the separation is dependent on stellar temperature, so that it decreases for colder faint-RGB stars. The same effect is also visible in the $C_{\rm BVI}$ index, which better identifies canonical/anomalous in bright RGB stars.
%Nevertheless, with anticipated improvements in the quality of Gaia XP spectra in forthcoming data releases, the performance of the $\rm f415^{25}$ filter is expected to enhance significantly. 
Despite these limitations, the newly defined $\rm f415^{25}$ filter successfully identifies canonical and anomalous stars. This effectiveness establishes a strong foundation for its future utilization in forthcoming data releases.

\subsection{Performance of the $\rm f360^{25}$ photometric filter}
\label{app:f360}

\begin{figure*}
    \centering
    \includegraphics[width=0.6\textwidth, trim={0cm 0cm 0cm 0cm}, clip]{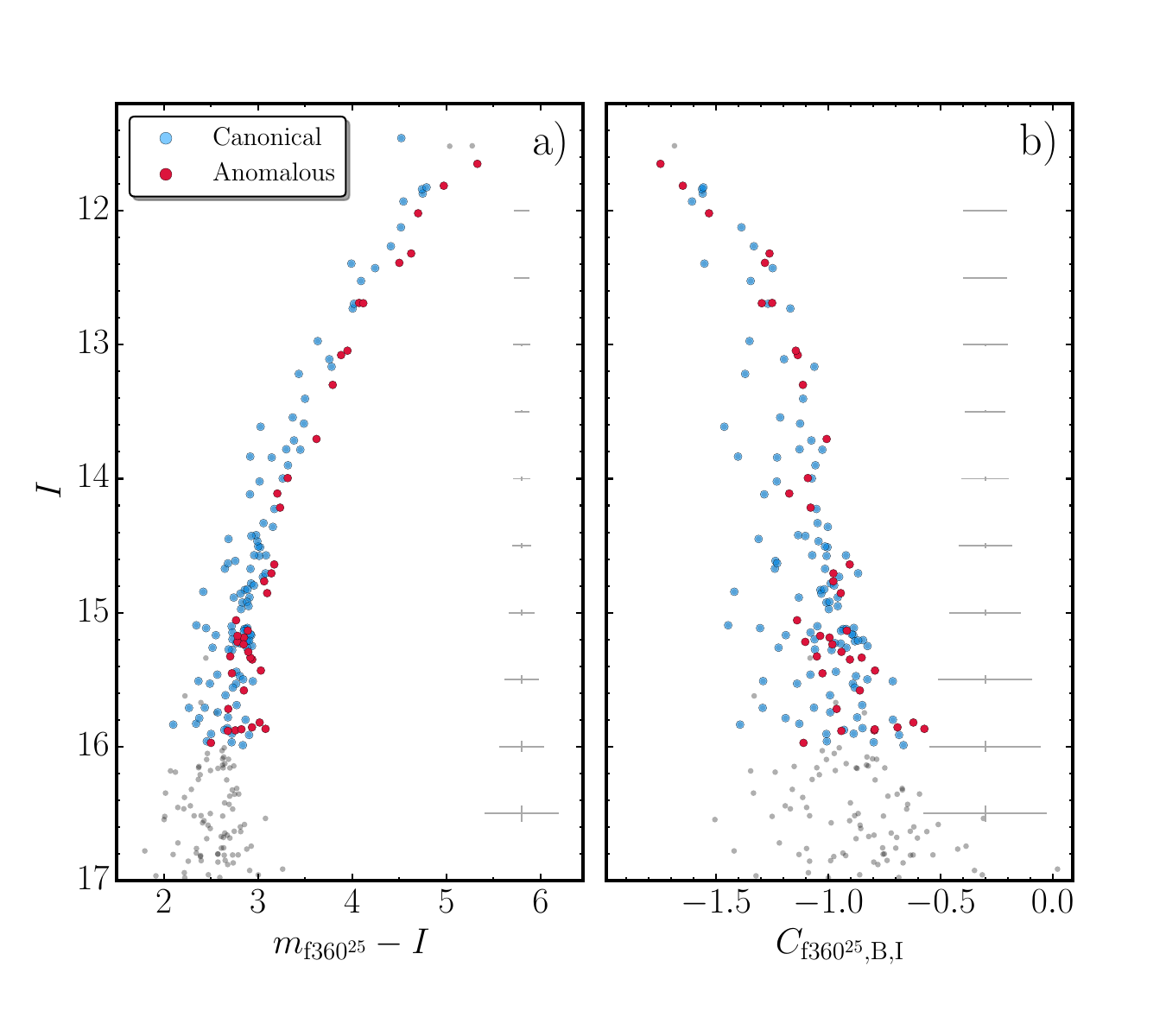}
    \caption{Performance of the $\rm f360^{25}$ photometric filter. \textit{Panel a).} $I$ vs. $m_{\rm f360^{25}}-I$ CMD. Azure and red markers indicate canonical and anomalous stars, while photometric uncertainties are shown with gray errorbars. \textit{Panel b).} $I$ vs. $C_{\rm f360^{25}, B, I} = (m_{\rm f360^{25}}-B)-(B-I)$ CMD. Color-coding is the same as in panel a).}
    \label{fig:f360}
\end{figure*}

Similar to the $\rm f415^{25}$ filter, the $\rm f360^{25}$ filter is defined as a box-filter encompassing the 350-375 nm wavelength range. Such filter is defined as to include the absorption feature located approximately between 350-390 nm \citep[see e.g. panels b1-b2 of Fig.~4 in][]{milone2017}, and to separate the bulk of 1P and 2P stars in NGC\,1851. The  $I$ vs. $m_{\rm f360^{25}}-I$ CMD, depicted in Fig.~\ref{fig:f360}a, suggests that the RGB-spread is larger than what is expected from observational uncertainties alone, even though the sequences are not distinctly separated enough to provide a reliable identification.
% To further investigate the performance of the $f360^{25}$ filers, we make use of the spectroscopic data from \citet{tautvaisiene2022} and examined the [Na/Fe] vs. [O/Fe] distribution of the stars in common with our dataset, shown in Fig.~\ref{fig:f360}b. Moreover, we selected Na-poor and Na-rich stars, i.e. 1P and 2P stars, as stars with [Na/Fe] smaller and larger than 0.2 dex, respectively. Na-poor stars (1P) are denoted by empty circles, while Na-rich stars (2P) are represented by crosses. Consistently with the color-coding adopted in the manuscript, azure and red markers indicate canonical and anomalous stars, respectively. The same scheme is applied for both panels of Fig.~\ref{fig:f360}.}
% \textbf{Upon a visual inspection of Fig.~\ref{fig:f360} we draw the following considerations, particularly when considering stars brighter than I=14, which exhibit lower photometric uncertainties (indicated by gray error bars in Fig.~\ref{fig:f360}a):
% i) Na-poor canonical stars (empty azure circles) are predominantly situated on the blue side of the CMD; ii) Na-rich canonical stars (azure crosses) are more widely distributed along the red giant branch (RGB);  iii) The anomalous population is almost exclusively composed of Na-rich stars, aligning with expectations from spectroscopic studies of NGC\,1851 \citep[see e.g. Fig.~11 of][]{tautvaisiene2022}, and they tend to lie on the red side of the RGB.
% Conversely, fainter stars, impacted by larger photometric uncertainties and flux fluctuations, tend to overlap regardless of their Na-poor/Na-rich canonical/anomalous classification.}
Nevertheless, a noticeable difference is observed in the RGB sequence of canonical and anomalous stars. Specifically, the RGB sequence of canonical stars spans a broader color range, whereas anomalous stars are predominantly located on a narrow sequence towards the red side of the RGB. This distinction becomes more prominent when considering stars brighter than $I=14$, as they exhibit lower photometric uncertainties (indicated by gray error bars in Fig.~\ref{fig:f360}). A plausible explanation for this behavior is that canonical stars display a wider range of light-element variations, in particular Nitrogen, compared to the anomalous stars \citep[see e.g.][]{lardo2012, yong2015, tautvaisiene2022}. Similar considerations are also true for Fig.~\ref{fig:f360}b, event though the photometric index is affected by slightly larger uncertainties.

Therefore, even though the $\rm f360^{25}$ filter yields promising results showing some indications of separation between 1P and 2P stars, it cannot be solely relied upon for unequivocal identification of first- and second-population stars in NGC\,1851. Nevertheless, as Gaia XP spectra continue to improve in future data releases, the $\rm f360^{25}$ filter may become a valuable tool for extracting essential information about multiple stellar populations at the clusters' outskirts.

\end{appendix}

%%%%%%%%%%%%%%%%%%%% REFERENCES %%%%%%%%%%%%%%%%%%

% The best way to enter references is to use BibTeX:

% Alternatively you could enter them by hand, like this:
% This method is tedious and prone to error if you have lots of references
%\begin{thebibliography}{99}
%\bibitem[\protect\citeauthoryear{Author}{2012}]{Author2012}
%Author A.~N., 2013, Journal of Improbable Astronomy, 1, 1
%\bibitem[\protect\citeauthoryear{Others}{2013}]{Others2013}
%Others S., 2012, Journal of Interesting Stuff, 17, 198
%\end{thebibliography}

%%%%%%%%%%%%%%%%%%%%%%%%%%%%%%%%%%%%%%%%%%%%%%%%%%

%%%%%%%%%%%%%%%%% APPENDICES %%%%%%%%%%%%%%%%%%%%%

% \appendix

% \section{Some extra material}

% If you want to present additional material which would interrupt the flow of the main paper,
% it can be placed in an Appendix which appears after the list of references.

%%%%%%%%%%%%%%%%%%%%%%%%%%%%%%%%%%%%%%%%%%%%%%%%%%

% Don't change these lines

\end{document}